\documentclass[twocolumn]{aastex63}
\usepackage{lineno}
\shorttitle{Are CL-AGNs Particular in Evolution Stage?}
\shortauthors{Wang et al.}

\begin{document}

\title{Are ``Changing-Look'' Active Galactic Nuclei Special in the Coevolution of 
Supermassive Black Holes and their Hosts? I.}

\correspondingauthor{J. Wang \& D. W. Xu}
\email{wj@nao.cas.cn, dwxu@nao.cas.cn}

\author{J. Wang}
\affiliation{Guangxi Key Laboratory for Relativistic Astrophysics, School of Physical Science and Technology, Guangxi University, Nanning 530004,
People's Republic of China}
\affiliation{Key Laboratory of Space Astronomy and Technology, National Astronomical Observatories, Chinese Academy of Sciences, Beijing 100101,
People's Republic of China}

\author{W. K. Zheng}
\affiliation{Department of Astronomy, University of California, Berkeley, CA 94720-3411, USA}

\author{T. G. Brink}
\affiliation{Department of Astronomy, University of California, Berkeley, CA 94720-3411, USA}

\author{D. W. Xu}
\affiliation{Key Laboratory of Space Astronomy and Technology, National Astronomical Observatories, Chinese Academy of Sciences, Beijing 100101,
People's Republic of China}
\affiliation{School of Astronomy and Space Science, University of Chinese Academy of Sciences, Beijing, People's Republic of China}

\author{A. V. Filippenko}
\affiliation{Department of Astronomy, University of California, Berkeley, CA 94720-3411, USA}

\author{C. Gao}
\affiliation{Guangxi Key Laboratory for Relativistic Astrophysics, School of Physical Science and Technology, Guangxi University, Nanning 530004,
People's Republic of China}
\affiliation{Key Laboratory of Space Astronomy and Technology, National Astronomical Observatories, Chinese Academy of Sciences, Beijing 100101,
People's Republic of China}
\affiliation{School of Astronomy and Space Science, University of Chinese Academy of Sciences, Beijing, People's Republic of China}

\author{C. H. Xie}
\affiliation{Guangxi Key Laboratory for Relativistic Astrophysics, School of Physical Science and Technology, Guangxi University, Nanning 530004,
People's Republic of China}
\affiliation{Key Laboratory of Space Astronomy and Technology, National Astronomical Observatories, Chinese Academy of Sciences, Beijing 100101,
People's Republic of China}
\affiliation{School of Astronomy and Space Science, University of Chinese Academy of Sciences, Beijing, People's Republic of China}

\author{J. Y. Wei}
\affiliation{Key Laboratory of Space Astronomy and Technology, National Astronomical Observatories, Chinese Academy of Sciences, Beijing 100101,
People's Republic of China}
\affiliation{School of Astronomy and Space Science, University of Chinese Academy of Sciences, Beijing, People's Republic of China}






\begin{abstract}

The nature of the so-called ``changing-look'' (CL) active galactic nucleus (AGN), which 
is characterized by spectral-type transitions within $\sim10$~yr, remains an open question. 
As the first in our series of studies, we here attempt to understand the CL phenomenon 
from a view of the coevolution of AGNs and their host galaxies (i.e., if CL-AGNs are at a specific 
evolutionary stage) by focusing on the SDSS local ``partially obscured'' AGNs in which the stellar 
population of the host galaxy can be easily measured in the integrated spectra. A spectroscopic follow-up program
using the Xinglong 2.16~m, Lick/Shane 3~m, and Keck 10~m telescopes enables us to identify 
in total 9 CL-AGNs from a sample of 59 candidates 
selected by their mid-infrared variability. 
Detailed analysis of these spectra shows that 
the host galaxies of the CL-AGNs are biased against young stellar populations and 
tend to be dominated by intermediate-age stellar populations. This motivates us to propose that 
CL-AGNs are probably particular AGNs at a specific evolutionary stage, such as
a transition stage from ``feast'' to ``famine'' fueling of the supermassive black hole.
In addition,  we reinforce the previous claim that CL-AGNs tend to be biased against both a high Eddington ratio and a high bolometric luminosity, suggesting that 
the disk-wind broad-line-region model is a plausible explanation of the CL phenomenon. 
\end{abstract}



\section{Introduction} \label{sec:intro}

``Changing-look'' active galactic nuclei (CL-AGNs), which are AGNs with a temporary
appearance or disappearance of their broad emission lines,
show a spectral transition between Type~1, intermediate type, and Type~2 within a timescale of years to decades (Ricci \& Trakhtenbrot 2022)\footnote{The CL-AGNs in this study are defined in optical band. 
In addition, CL-AGNs include X-ray
objects with a significant variation of line-of-sight column density (e.g., Risaliti et al. 2009; Marinucci et al. 2016).}.
The CL phenomenon is rare; thus far,
with multi-epoch photometry and optical spectroscopy, only $\sim 150$ CL-AGNs have been identified
(e.g., MacLeod et al. 2010, 2016, 2019; Shapovalova et al. 2010; Shappee et al. 2014; LaMassa et al. 2015; McElroy et al. 2016; Parker et al. 2016;
Ruan et al. 2016; Runnoe et al. 2016; Gezari et al. 2017; Sheng et al. 2017, 2020; Kollatschny et al. 2018, 2020; Stern et al. 2018;
Wang et al. 2018, 2019, 2020a, 2022; Yang et al. 2018; Frederick et al. 2019; Guo et al. 2019; Trakhtenbrot et al. 2019; Yan et al. 2019;
Ai et al. 2020; Graham et al. 2020; Nagoshi et al. 2021;  
Green et al. 2022; Hon et al. 2022; Marin et al. 2019; Parker et al. 2019; Mathur et al. 2018; L{\'o}pez-Navas et al. 2022, 2023).

Recently, many studies have been carried out of CL-AGNs (e.g., Nagoshi \& Iwamuro 2022;
Panda \& Sniegowska 2022, and references above) owing to their peculiarity.
The CL phenomenon shows that the different spectral types cannot be fully understood by
the widely accepted orientation-based AGN unified model (e.g., Antonucci 1993), in which
the central engine is obscured in Type~2 AGNs by the dusty torus along the line of sight to an observer. Moreover, it 
challenges the standard disk model in terms of the viscosity crisis
(e.g., Lawrence 2018, and references therein).

Currently, the physical origin of CL-AGNs is an open question. Several interpretations have been proposed,
including an accelerating outflow (e.g., Shapovalova et al. 2010), a variation of the obscuration (e.g., Elitzur 2012)\footnote{In addition to the objects with an appearance and/or disappearance of broad emission lines,
CL-AGNs include X-ray objects with a significant variation of line-of-sight column density
(e.g., Risaliti et al. 2009; Marinucci et al. 2016).}, a tidal disruption event (e.g., Merloni et al. 2015; Blanchard et al. 2017), and an accretion-rate change (e.g., Elitzur et al. 2014; Gezari et al. 2017; Sheng et al. 2017; Yang et al. 2018; Wang et al. 2018, 2019, 2020a,b, 2022; Guo et al. 2019).
There is, in fact, accumulating evidence supporting the scenario that the CL phenomenon results from a variation of accretion power of a supermassive black hole (SMBH; e.g., Feng et al. 2021a), even though the
physics behind the CL phenomenon is still poorly understood (e.g., Saade et al. 2022; Ren et al. 2022).
Moreover, CL-AGNs provide us with an ideal opportunity to examine the coevolution between AGNs and their host galaxies (e.g., Heckman \& Best 2014) 
by directly investigating the host-galaxy properties of luminous AGNs. 

However, the properties of the host galaxies of CL-AGNs are still actively debated,  depending on the adopted sample and analysis methods.  
On the one hand, some studies have
shown that CL-AGNs and non-CL-AGNs (NCL-AGNs) share similar host-galaxy properties (e.g., Charlton et al. 2019; Yu et al. 2020; Dodd et al. 2021).
On the other hand,  a special stellar population has been reported for CL-AGNs by some recent studies (e.g., Liu et al. 2021; Jin et al. 2022).
Our current understanding of the host galaxies of CL-AGNs is greatly hindered by two facts: (1) the sample size of confirmed CL-AGNs is small, and 
(2) starlight from the host is usually overwhelmed by strong nonstellar radiation from the luminous active nucleus. 
To address the second issue, we here report a pilot study of CL-AGNs in the 
context of the coevolution between 
SMBHs and their hosts by focusing on the CL-AGNs newly 
identified from a sample of local ``partially obscured'' AGNs, whose optical spectra 
enable us to measure the circumnuclear stellar populations of the host galaxies. 

The paper is organized as follows.
Section 2 presents the sample selection.
Optical spectroscopy and X-ray follow-up observations,
along with data reduction, are described in Section 3. 
Sections 4 presents our CL-AGN identification. The spectral analysis and statistical results 
are given in Section 5, and we discuss our conclusions in Section 6. 
A $\Lambda$CDM cosmological model with parameters H$_0=70\,\mathrm{km\,s^{-1}\,Mpc^{-1}}$, $\Omega_{\mathrm{m}}=0.3$, and
$\Omega_\Lambda=0.7$ is adopted throughout.

\section{Sample Selection} \label{sec:style}

We start from the SDSS local (redshift $z=0.011$--0.025) ``partially obscured'' AGNs studied comprehensively 
by Wang (2015). The sample in total contains 170 broad-line Seyfert galaxies/LINERs and
44 broad-line composite galaxies, all with high-quality SDSS spectra. A subsample 
of CL-AGN candidates are then selected from the 170 broad-line Seyfert galaxies/LINERs
through two steps. 

Initially, the 170 objects are crossmatched with the catalog of the 
{\it Wide-field Infrared Survey Explorer} ({\it WISE} and {\it NEOWISE-R}; Wright et al. 2010; 
Mainzer et al. 2014). 
Based on the study by Sheng et al. (2020),
we exclude 84 objects with stable mid-infrared (MIR) light 
curves in both the $w1$ (3.4~$\mu$m) and $w2$ (4.6~$\mu$m) bands as follows 
At first, each light curve was smoothed
by an averaged measurement, along with the corresponding uncertainty,
within one day. Second, 45 light curves are in total identified as a 
stable one with $\mathrm{std}/\sigma_{\mathrm{mode}}<1$. $\mathrm{std}$ is the 
standard deviation calculated from all the averaged measurements, and 
$\sigma_{\mathrm{mode}}=3\times\sigma_{\mathrm{median}}-2\times\sigma_{\mathrm{mean}}$
the mode of the uncertainty of individual averaged measurement, 
where $\sigma_{\mathrm{median}}$ and $\sigma_{\mathrm{mean}}$ are the corresponding median and mean 
values, respectively. Finally, among the remaining 125 light curves, 39 objects were additionally
excluded by visual inspection, because of their light curves with either strong fluctuation or 
unexpected large deviation. The previous studies (see citations in Section 1) in fact show that 
CL-AGNs are typical of smooth and long term ($\sim1-10$ yrs) variability in MIR. 

Second, the SDSS $g$-band brightness is required to be no fainter than 17.5~mag so that optical 
spectra of adequate quality with a median signal-to-noise ratio (S/N) per pixel of the 
whole spectrum $\mathrm{S/N>20}$ could be obtained by the Xinglong 2.16m and Lick/Shane 3m telescopes 
(see below )within an exposure time less than 3600 seconds.
This left 59 CL-AGN candidates for subsequent spectroscopy.

\section{Observations and Data Reduction}

\subsection{Optical Spectroscopy}
\subsubsection{Observations}

The follow-up long-slit spectroscopy of the 59 CL-AGN candidates 
was carried out by 
the Beijing Faint Object Spectrograph and Camera (BFOSC) mounted 
on the 2.16~m telescope (Fan et al. 2016) at the Xinglong Observatory
of the National Astronomical Observatories, Chinese Academy
of Sciences (NAOC), and by the Kast double spectrograph (Miller \& Stone 1994) 
mounted on the Shane 3~m telescope at Lick Observatory.  
In practice, we at first excluded the objects without a spectral variation (continuum shape 
and emission lines) by the spectra taken by the 2.16 telescope. 
In order to identify CL phenomenon by the spectral analysis and method described in Section 5,
spectra of higher quality were then obtained for the remaining objects by the Lick/Shane telescope. 

The spectra taken by the 2.16~m telescope were
obtained with the G4 grism and a long slit of width 2\arcsec\ oriented in the
north–south direction, leading to a spectral resolution of $\sim 10$~\AA\ 
and a wavelength coverage of 3850–-8200~\AA. Wavelength calibration was 
carried out with spectra of iron-argon comparison lamps. In order to minimize
the effects of atmospheric dispersion (Filippenko 1982), all
spectra were obtained as close to the meridian as possible.

The Kast spectra were obtained with the 2\arcsec-wide slit,
the 600/4310 grism on the blue side, and the 300/7500 grating on the red side.  This configuration
produced wavelength resolutions of $\sim 5$~\AA\ and $\sim 12$~\AA\ on the blue and red sides (respectively), and a combined wavelength range of 3600--10,700~\AA.
The slit was aligned at or near the parallactic angle (Filippenko 1982) to minimize
differential light losses caused by atmospheric dispersion.

SDSS\,J151652.48+395413.4 was additionally observed with the DEep Imaging Multi-Object Spectrograph (DEIMOS; Faber et al. 2003) mounted on the Keck~II 10~m telescope on 12 May 2023 (UTC dates are used throughout this paper). The 600~line~mm$^{-1}$ grating and a 1\arcsec-wide slit were utilized, resulting in a spectral resolution of 5~\AA\ and a wavelength range of 4500--9600~\AA.

All of the spectra were flux calibrated with observations of Kitt Peak National Observatory standard stars (Massey et al. 1988).
The exposure time for each object ranges from 300~s to 3600~s, depending on the telescope size,  object brightness, and weather conditions.

\subsubsection{Data Reduction}

One-dimensional (1D) spectra were extracted from the raw
images by utilizing IRAF\footnote{IRAF is distributed by NOAO, which is operated by AURA, Inc., under cooperative agreement with the U.S. National Science Foundation (NSF).} (Tody 1986, 1992)
packages and standard procedures for bias subtraction and flat-field
correction.

All of the extracted 1D spectra were then calibrated in
wavelength and in flux with spectra of the comparison lamps
and standard stars. The accuracy of the wavelength calibration is
better than 1~\AA\ for the Kast and DEIMOS spectra, and better than 2~\AA\ for the
BFOSC spectra. The telluric A-band (7600–-7630~\AA) and B-band (around 6860~\AA) absorption produced by atmospheric $\mathrm{O_2}$
molecules were removed from the extracted spectra by using the standard-star spectra.
Each calibrated spectrum was then corrected for Galactic
extinction according to the color excess
$E(B-V)$ taken from the Schlegel \& Finkbeiner (2011) Galactic
reddening map.
The correction was applied by assuming the $R_V = 3.1$ extinction law of our 
Galaxy (Cardelli et al. 1989).
Spectra were then transformed to the rest frame according to their redshifts. 

\subsection{X-ray Observations and Data Reduction}

X-ray follow-up observations of two
objects (the newly identified CL-AGNs at their ``turn-off'' states; see Section 4), 
SDSS\,J124610.75+275615.9 and SDSS\,J151652.48+395413.4, were carried out by 
using the {\it Neil Gehrels Swift Observatory} (Gehrels et al. 2004) X-ray telescope
(XRT). Both objects were targeted (ObsID = 00016002001 and 00016004001)
on 2023 May 03. The exposure times were
1281~s and 1636~s for SDSS\,J124610.75+275615.9 and SDSS\,J151652.48+395413.4, respectively,
in the XRT Photon Counting (PC) mode.

We reduced the XRT data by 
HEASOFT version 6.27.2, along with the corresponding
CALDB version 20190910. For each of the two objects, the source spectrum was
extracted from the image in a circular region with a radius of
10.0\arcsec. An adjacent region free of any sources was adopted to
extract the background-sky spectrum.
The task \it xrtmkarf \rm was used to generate the corresponding
ancillary response file.

\section{Identification of New CL-AGNs}

For each of the 59 ``partially obscured'' CL-AGN candidates, a differential spectrum in the rest frame
is created from a pair of the Shane\footnote{We use the Shane spectra in the 
subsequent spectral identification and analysis, simply because of their higher 
S/N when compared with the BEFOSC spectra.}
and SDSS DR16 spectra, and is used to search for new CL-AGNs. 
Each differential spectrum is created as follows. 
First, the SDSS DR16 spectrum is 
convolved with a Gaussian function with a velocity dispersion $\sigma$ to match the spectral resolution of the 
corresponding Shane spectrum through $\sigma=\sqrt{\sigma_{\mathrm{Shane}}^2-\sigma_{\mathrm{SDSS}}^2}$, where $\sigma_{\mathrm{Shane}}$ and $\sigma_{\mathrm{SDSS}}$
are the instrumental resolution of the Shane and SDSS DR16 spectra, respectively.
A zero-point correction of the wavelength calibration is then applied to each Shane spectrum 
by matching the [\ion{O}{3}] $\lambda5007$ line centroid in the rest frame. 
After scaling the flux level by the [\ion{O}{3}] $\lambda5007$ line flux, 
a differential spectrum is built by $\Delta f_{\lambda}=f_{\lambda,\mathrm{Shane}}-f_{\lambda,\mathrm{DR16}}$.

With the differential spectra, we at first extract 12 sources with an evident line variation by
requiring the signal-to-noise ratio of 
flux variation $\Delta F/\sigma_l>5.0$ for either H$\beta$ or H$\alpha$ emission lines, 
where $\Delta F$ is the flux integrated over a proper wavelength range\footnote{The wavelength 
range is determined in advance according to our previous study in Wang (2015).} 
on each differential spectrum.
$\sigma_l$ is the corresponding statistic error of the line flux, 
which is determined by the method given in 
Perez-Montero \& Diaz (2013)\footnote{$\sigma_l=\sigma_cN[1+EW(N\Delta)]$, where $\sigma_c$ is the standard deviation of continuum in a box near the line, $N$ the number of pixels used to 
measure the line flux, $EW$ the equivalent width of the line, and $\Delta$ the wavelength 
dispersion in units of $\mathrm{\AA\ pixel^{-1}}$.}. 

With an examining the differential spectra individually by eye and the spectral analysis 
(see Section 5) which enables us to determine an existence of broad 
Balmer emission line, 9 new CL-AGNs are in total identified by either of the two criteria: 
(1) a disappearance of 
broad H$\alpha$ emission in our new spectra; (2) an appearance/disappearance of broad 
H$\beta$ emission in our new spectra if the broad H$\beta$ was not/was required in the SDSS spectra.  
Among the 9 new CL-AGNs, SDSS\,J075244.20+455657.4 (B3\,0749+460A) has been reported and studied separately by Wang et al. (2022);
see that paper for the details of the data reduction and spectral analysis. 
A log of the eight new CL-AGNs identified here is listed in Table 1. 

The objects are displayed in 
Figure 1 not only by a comparison between the Shane and SDSS DR16 spectra, but also by 
the corresponding differential spectrum. 
One can see that the 
``turn-on'' fraction resulting from our spectroscopic campaign is 5/9. This high fraction is 
not hard to understand, since the ``partially obscured'' AGNs typically have a 
Seyfert 1.5 to 1.9-like spectrum that is usually far from a ``turn-on'' state.

\begin{figure*}[htp!]
\plotone{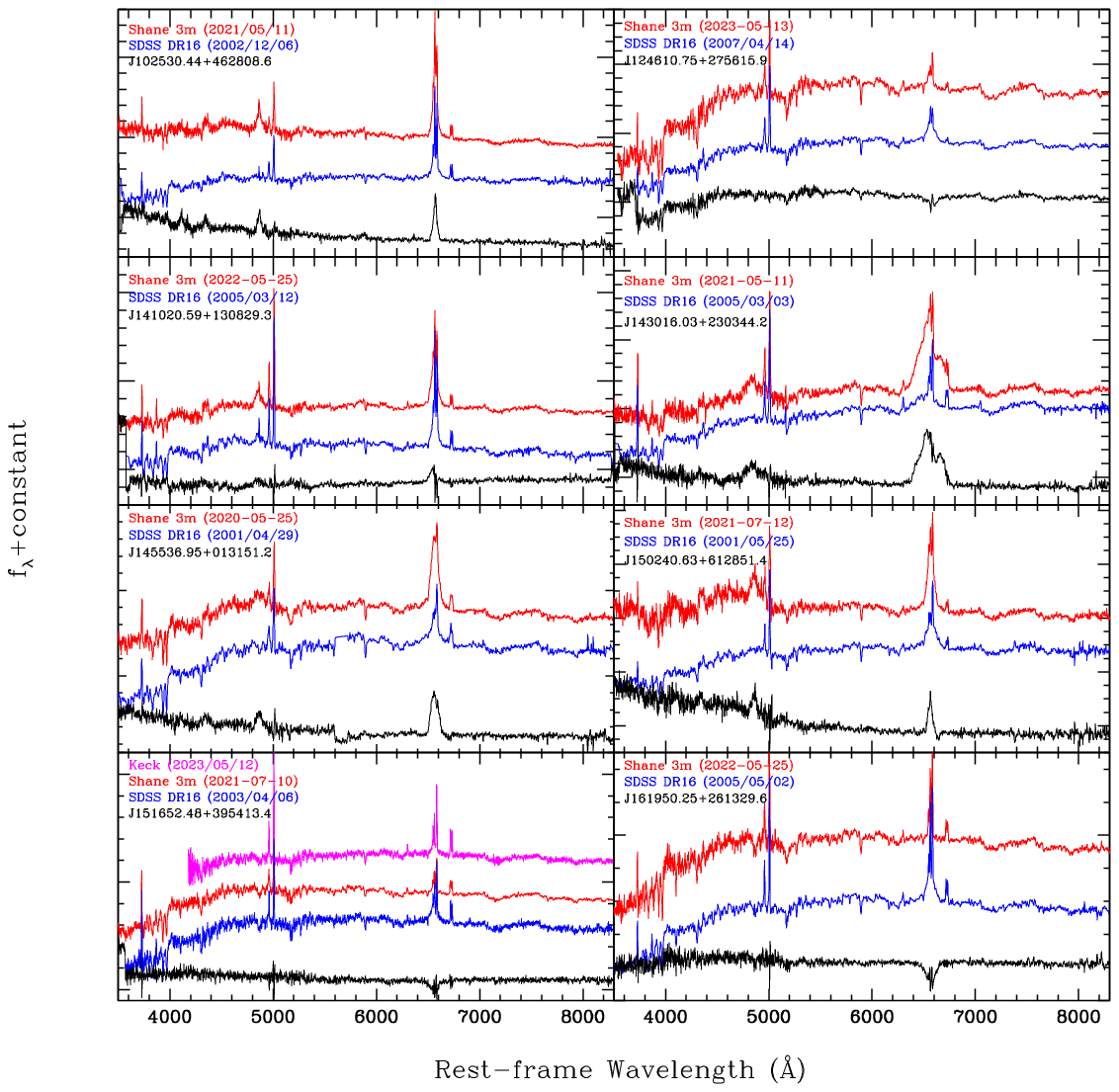}
\caption{A comparison of the multi-epoch rest-frame spectra of the 8 newly identified
CL-AGNs.  In each panel, the Shane and SDSS DR16 spectra are denoted 
by the red and blue lines, respectively.  The bottom black lines are the differential spectrum, when the
SDSS DR16 spectrum is used as a reference. Before the subtraction, the two spectra are 
matched in the instrumental resolution, and then scaled according to 
their [\ion{O}{3}] $\lambda$5007 line fluxes. The spectra are shifted vertically for clarity. 
\label{fig:general}}
\end{figure*}

\section{Analysis and Results}

In order to quantify the CL phenomena identified in the eight new CL-AGNs and to reveal the underlying physics, 
spectral analysis was performed by following our previous studies (e.g., Wang 2015; 
Wang et al. 2019, and references therein).

\subsection{Continuum Removal}

As shown in the Figure 1, 
the continuum of the CL ``partially obscured'' AGNs is dominated by
the starlight component emitted from the host galaxies, even in the ``turn-on'' state.
We therefore model the stellar absorption features in each Shane spectrum by the method
adopted by Wang (2015). Briefly, to isolate the  emission-line spectrum, 
the continuum of each Shane spectrum is fitted by a linear combination of the 
first seven eigenspectra that are built
through the principal-component analysis (PCA) method (e.g.,
Francis et al. 1992; Hao et al. 2005; Wang \& Wei, 2008) from
the standard single stellar population spectral library developed
by Bruzual \& Charlot (2003). 
In addition, an intrinsic extinction due to the 
host galaxy described by a Galactic extinction curve with $R_V=3.1$ is involved 
in our continuum modeling.

For each spectrum, a $\chi^2$ minimization is performed iteratively over the entire spectral wavelength range, except for the regions with
known strong emission lines, such as low-order Balmer lines (both
narrow and broad components), [\ion{S}{2}] $\lambda\lambda$6716, 6731,
[\ion{N}{2}] $\lambda\lambda$6548, 6583, [\ion{O}{1}] $\lambda$6300, [\ion{O}{3}] $\lambda\lambda$4959, 5007, [\ion{O}{2}] $\lambda\lambda$3726, 3729,
[\ion{Ne}{3}] $\lambda$3869, and [\ion{Ne}{5}] $\lambda$3426.
In the minimization, the velocity dispersion of the stellar component is a free parameter for
the SDSS DR16 spectra, and  
is fixed in advance for the Shane/Kast spectrum, because the observed absorption features are dominated by the instrumental profile. As an example, the subtraction of the starlight component is illustrated in the left panels of Figure 2 
for SDSS\,J102530.44+462808.6.

\begin{figure*}[ht!]
\plotone{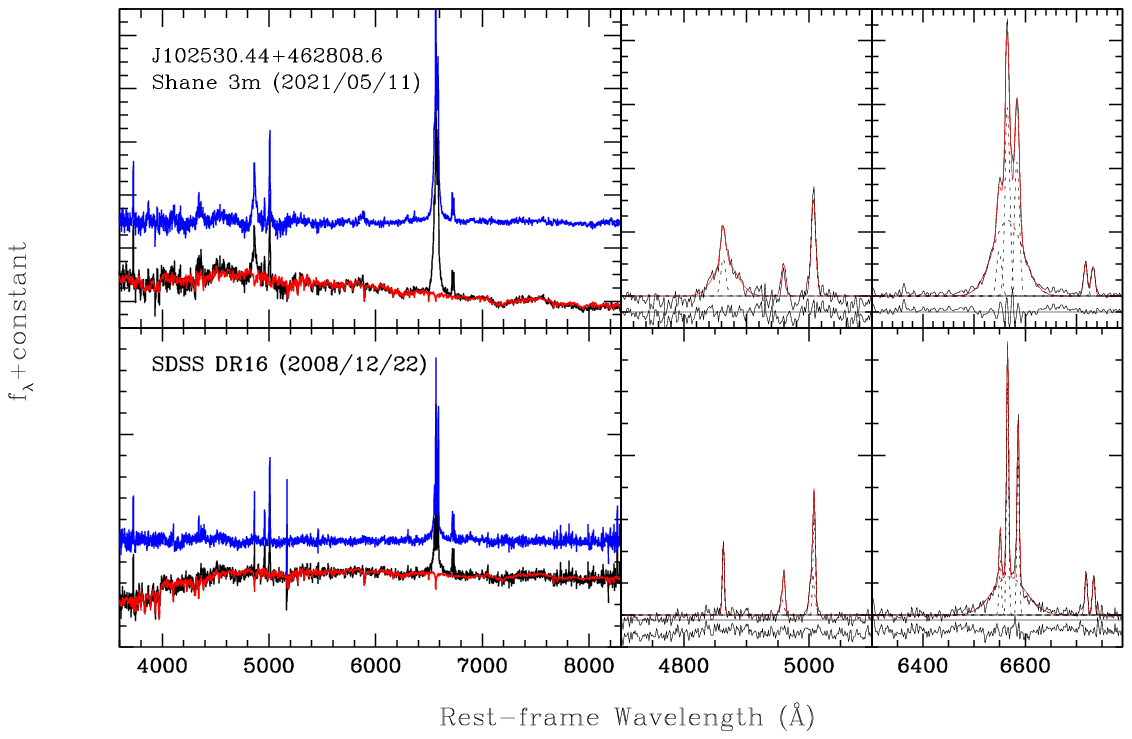}
\caption{Illustration of analysis for the multi-epoch spectra of SDSS\,J102530.44+462808.6.  
{\it Left panels:} modeling and removal of the continuum by a linear combination of 
the first seven eigenspectra that are extracted from 
the standard single stellar population spectral library developed
by Bruzual \& Charlot (2003).
In each panel, the top-blue curve shows the continuum-removed emission-line spectrum. The heavy black curve underneath shows 
the observed rest-frame spectrum, overplotted by the best-fit continuum indicated by the red curve.  
{\it Middle panels:} an illustration of the line-profile modeling with a linear combination of a set of Gaussian functions
for the H$\beta$ region. In each panel, the modeled continuum has already
been removed from the originally observed spectrum. The observed and modeled line profiles are plotted
with black and red solid lines, respectively. Each used Gaussian function is shown with a dashed line.
The subpanel underneath the line spectrum presents the residuals between the observed and modeled
profiles. {\it Right panels:} same as the middle panels, but for the H$\alpha$ region.
\label{fig:general}}
\end{figure*}

\subsection{Line-Profile Modeling}

After removing the underlying continuum in each of the spectra,
a linear combination of a set of Gaussians is adopted to 
model the emission-line profiles in both the H$\alpha$ and H$\beta$ regions by the SPECFIT task (Kriss 1994) in IRAF.
In the modeling, the line-flux ratios of the [\ion{O}{3}] $\lambda\lambda$4959, 5007 and [\ion{N}{2}] $\lambda\lambda$6548, 6583 doublets are 
fixed to their theoretical values of 1:3 (e.g., Dimitrijevic et al. 2007). In addition to a narrow component, a blueshifted broad component is necessary for 
reproducing the [\ion{O}{3}] $\lambda\lambda$4959, 5007 line profiles in a fraction of the spectra 
(e.g., Boroson 2005; Zhang et al. 2013; Harrison et al. 2014; Woo et al. 2017; Wang et al. 2011, 2018). 
As an example,
the line-profile modeling is illustrated in the middle and right panels of Figure 2 for 
the H$\beta$ and H$\alpha$ regions, respectively.

The results of our spectral analysis are given in Table 1. 
The redshift and UTC observation date are listed in Columns (2) and (3), respectively. 
Columns (4), (5), and (6) give the measured line fluxes of [\ion{O}{3}] $\lambda$5007, broad H$\beta$, 
and broad H$\alpha$ emission, respectively. 
The widths of the broad H$\beta$ and H$\alpha$ are shown in Columns (7) and (8), respectively.
In SDSS\,J143016.03+230344.2, the H$\alpha$ broad emission has to be reproduced by two
Gaussian functions. The full width at half-maximum intensity (FWHM)
of the integrated broad-line emission is 
measured from a residual profile that is obtained by subtracting the modeled narrow component from the observed profile.
In addition, the corresponding CL phenomenon status (``turn on'' or ``turn off'') is 
shown in Column (12).
The ``turn-off'' state corresponds to undetectable 
broad H$\beta$ or H$\alpha$ emission, and the ``turn-on'' state has easily evident broad H$\beta$ emission.

All of the uncertainties reported in Table 1 correspond to the 1$\sigma$ significance level and include only the uncertainties caused by the fitting,
rather than the removal of the stellar continuum.

\begin{longrotatetable}
\begin{deluxetable*}{ccccccccccccc}
\tablecaption{Results of Line-Profile Modeling and Analysis\label{chartable}}
\tabletypesize{\tiny}
\tablehead{
\colhead{SDSS ID} &
\colhead{$z$}  &
\colhead{UT Date} & 
\colhead{$F(\mathrm{[O~III]} \lambda5007)$} &
\colhead{$F(\mathrm{H\beta_b})$} & 
\colhead{$F(\mathrm{H\alpha_{b}})$} &
\colhead{($\mathrm{FWHM\beta_b}$)} & 
\colhead{$\mathrm{FWHM(H\alpha_{b})}$}  & 
\colhead{$\log(M_{\mathrm{BH}}/{\rm M}_\odot)$} &
\colhead{$L/L_{\mathrm{Edd}}$} & 
\colhead{$D_{\mathrm{n}}(4000)$} &
\colhead{H$\delta_{\mathrm{A}}$} &
\colhead{Status}\\
\colhead{} & \colhead{}  & \colhead{}  &
\multicolumn{3}{c}{$\mathrm{(10^{-15}\ erg\ s^{-1}\ cm^{-2})}$} &
\multicolumn{2}{c}{$\mathrm{(km\ s^{-1})}$} & & & & (\AA) &\\
\cline{4-5} 
\cline{7-8}
}
\colnumbers
\startdata
J102530.44+462808.6 &  0.0795948 & 2002-12-06 & $2.70\pm0.26$ & \dotfill &  $9.95\pm0.24$  & \dotfill & $3580\pm130$ &  7.36 & 0.020 & 1.39 & 2.39 & turn off \\
\dotfill            & \dotfill   & 2021-05-11 & $2.91\pm0.12$ & $5.02\pm0.35$ & $17.48\pm0.59$ & $2770\pm230$ & $3160\pm110$ & 7.48 & 0.048 & \dotfill & \dotfill & turn on\\ 
J124610.75+275615.9 & 0.0230846  & 2007-04-14 & $13.53\pm0.15$ & \dotfill & $24.89\pm0.53$ & \dotfill & $5430\pm170$ & 7.33 & 0.003 & 1.83 & -0.75 &  turn on \\
\dotfill            & \dotfill   & 2021-07-12 & $15.34\pm0.57$ & \dotfill & \dotfill & \dotfill & \dotfill & 
\dotfill & \dotfill & \dotfill & \dotfill & turn off\\
\dotfill            & \dotfill   & 2023-05-13 & $8.40\pm0.17$ & \dotfill & \dotfill & \dotfill & \dotfill & 
\dotfill & 0.001\tablenotemark{a}  & \dotfill & \dotfill & turn off\\
J141020.59+130829.3 & 0.0593172 & 2005-03-12 &  $6.34\pm0.55$  & \dotfill & $12.23\pm0.23$ & \dotfill & $3310\pm70$ & 7.10 &  0.016 & 1.38 & -0.02 & turn off\\
\dotfill            & \dotfill  & 2022-05-25 &  $9.00\pm0.18$  & $6.35\pm0.30$ & $3690\pm210$ & $26.76\pm0.40$ & $3600\pm60$ &  7.42 &  0.026 & \dotfill & \dotfill & turn on \\
J143016.03+230344.2 & 0.0809661 & 2005-03-03 &  $8.49\pm0.34$  & \dotfill & $17.56\pm0.54$ & \dotfill & $9350\pm320$ & 8.43 &  0.004 & 1.61 & -0.48 & turn off \\
\dotfill            & \dotfill  & 2021-05-11 &  $8.00\pm0.21$  & $12.55\pm0.72$ & $9580\pm640$ & $85.29\pm0.63$ & $3360\pm160$ & 7.84 &  0.08 & \dotfill & \dotfill & turn on \\
J145536.95+013151.2 & 0.0973581 & 2001-04-29 &  $4.92\pm0.52$  & \dotfill & $12.83\pm0.41$ & \dotfill & $7760\pm320$ & 8.34 & 0.008 & 1.59 & 0.32 & turn off\\
\dotfill            & \dotfill  & 2020-05-25 &  $7.11\pm0.28$  & $9.49\pm0.67$ & $52.19\pm0.73$ & $4650\pm430$ & $4380\pm70$ & 7.73 &  0.024 & \dotfill & \dotfill & turn on \\
J150240.63+612851.4 & 0.109168 & 2001-05-25 &  $3.39\pm0.09$   & \dotfill & $12.39\pm0.23$ & \dotfill & $5660\pm130$ & 8.43 & 0.004 &  1.52 & -1.56 & turn off \\
\dotfill            & \dotfill & 2021-07-12 &  $4.50\pm0.22$   & $8.37\pm0.60$ & $24.17\pm0.41$ & $4340\pm120$ & $4290\pm100$ & 7.84 &  0.081 & \dotfill & \dotfill & turn on \\
J151652.48+395413.4 & 0.0632311 & 2003-04-08 & $5.62\pm0.09$   & \dotfill  & $11.34\pm0.32$ & \dotfill & $4350\pm170$ &  7.57 &  0.014 & 1.66 & 1.84 & turn on\\
\dotfill            & \dotfill & 2021-07-10  & $7.26\pm0.18$   & \dotfill  & \dotfill & \dotfill & \dotfill & 
\dotfill & \dotfill & \dotfill & \dotfill & turn off\\
\dotfill            & \dotfill & 2023-05-12  & $5.38\pm0.06$   & \dotfill  & $5.28\pm0.25$ & \dotfill & $4670\pm400$ & 
7.43 & 0.010/0.027\tablenotemark{a} & \dotfill & \dotfill & turn on\\
J161950.25+261329.6 & 0.0955506 & 2005-05-02 & $6.30\pm0.09$   & \dotfill  & $15.95\pm0.35$  & \dotfill & $4890\pm140$ &  7.84 &  0.016 & 1.48 & 3.01 & turn on \\
\dotfill            & \dotfill & 2022-05-25 &  $6.36\pm0.22$   & \dotfill  & \dotfill & \dotfill & \dotfill & 
\dotfill & \dotfill & \dotfill & \dotfill & turn off\\
 \enddata
\tablenotetext{a}{The values are estimated from the unabsorbed X-ray luminosity in the 2--10~keV energy band.}
\end{deluxetable*}
\end{longrotatetable}

\subsection{X-Ray Emission}

The total XRT count rates in the 0.3-–10~keV range were 
$(1.11\pm0.30)\times10^{-2}$ and $(1.04\pm0.26)\times10^{-2}\ \mathrm{count\ s^{-1}}$
for SDSS\,J124610.75+275615.9 and SDSS\,J151652.48+395413.4, respectively.
Owing to their low count rates, the X-ray energy spectra of both objects 
are modeled by XSPEC (v12.11; Arnaud 1996) with
a simple model of $wabs*zwabs*powerlaw$ over the 0.3-–10~keV range in terms of
the C-statistic (Cash 1979; Humphrey et al. 2009; Kaastra 2017). In the fitting,
the power-law photon index is fixed to be 2, and  
the Galactic hydrogen column density values are taken from 
the Leiden/Argentine/Bonn (LAB) Survey (Kalberla et al. 2005). 
The best fits are displayed in Figure 3.
The modeled hard X-ray flux in 2--10~keV energy band $F_{\mathrm{2-10~keV}}$, along with a comparison with previous measurements,
are given in Table 2.

One can see from the table a decreased $F_{\mathrm{2-10~keV}}$ for the ``turn-off'' state of SDSS\,J124610.75+275615.9.
In SDSS\,J151652.48+395413.4, the value of $F_{\mathrm{2-10~keV}}$ is estimated to be
comparable to that reported in the {\it ROSAT} catalog (Zimmermann et al. 2001; Voges et al. 1999), which agrees with the observed weak H$\alpha$ broad emission.    

\begin{figure}[ht!]
\plotone{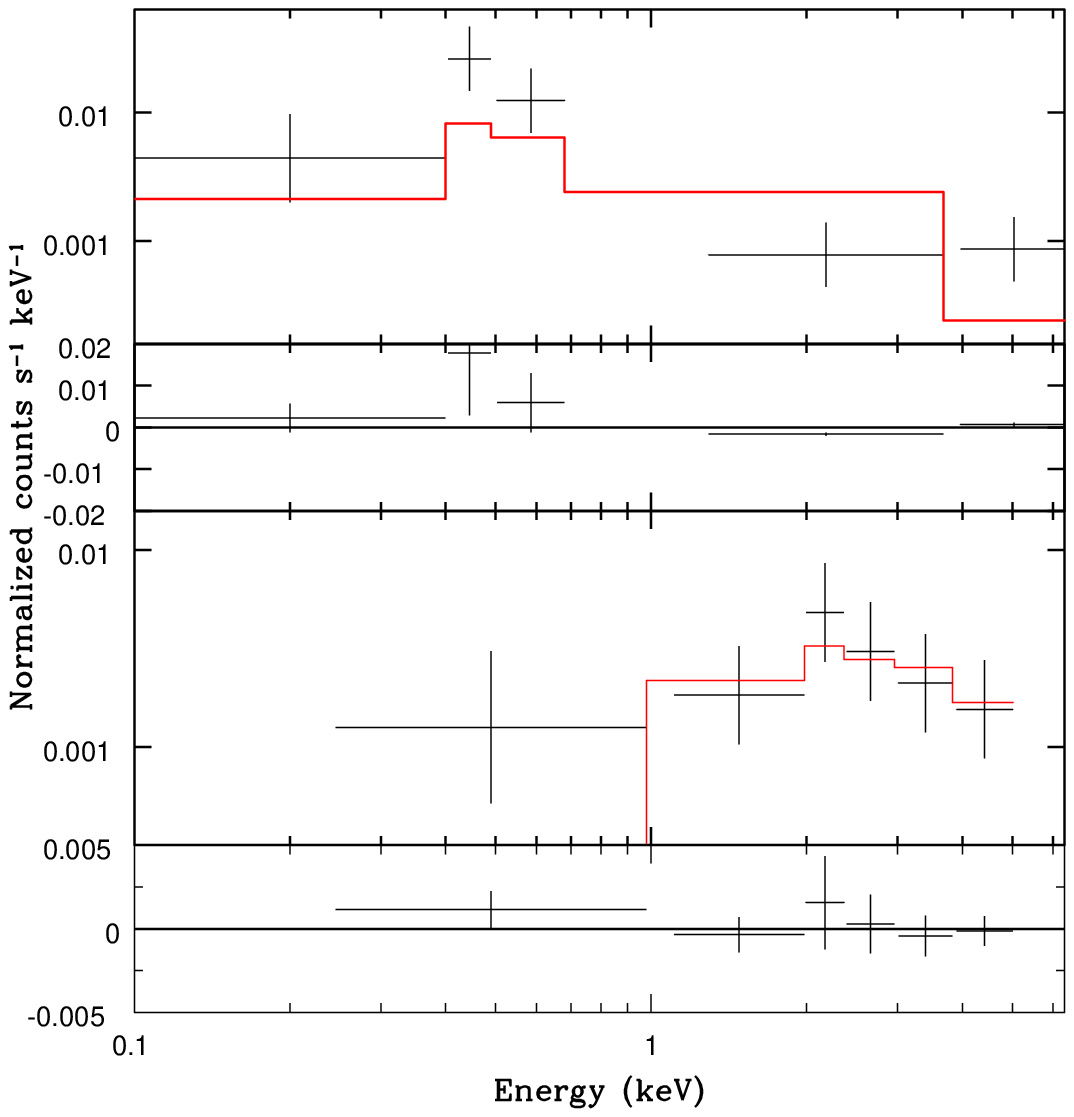}
\caption{{\it Swift}/XRT X-ray spectra of SDSS\,J124610.75+275615.9 (upper panel) and SDSS\,J151652.48+395413.4 (lower panel), both taken on 2023-05-03.
The best-fit spectral models expressed as $wabs*zwabs*powerlaw$ with a fixed power-law photon index of 2 are overplotted by the red solid lines.
The subpanel underneath each spectrum shows the deviations, in units of $\mathrm{counts\ s^{-1}\ keV^{-1}}$ , of the
observed data from the best-fit model.
\label{fig:general}}
\end{figure}

\begin{table*}
        \centering
        \caption{Hard X-ray flux in the 2--10~keV energy band.}
        \label{tab:example_table}
        \begin{tabular}{ccccc} 
                \hline
                \hline
                   Object  & Mission &  Date & $F_{\mathrm{2-10~keV}}$ & Reference\\
                              &             &          &   ($\mathrm{erg\ s^{-1}\ cm^{-2}}$) & \\
                         (1)  &   (2)  & (3) & (4) & (5) \\  
                         \hline 
                   SDSS\,J124610.75+275615.9 & {\it ROSAT} & 1990-1991 & $(1.9\pm0.4)\times10^{-13}$ & Zimmermann et al. (2001)\\
                                                            & Swift & 2023-05-03 & $(1.6\pm0.9)\times10^{-13}$ & This work\\
                   SDSS\,J151652.48+395413.4 & XMM-Newton& 2014-08-13 & $(2.3\pm1.2)\times10^{-12}$ & XMM-SSC (2018) \\
                                                            & Swift & 2023-05-03 & $(9.9\pm6.7)\times10^{-13}$ & This work\\
                  \hline
        \end{tabular}
\end{table*}

\subsection{Estimation of Black Hole Mass and Eddington Ratio}

Columns (9) and (10) in Table 1 list the estimated $M_{\mathrm{BH}}$ and $L_{\mathrm{bol}}/L_{\mathrm{Edd}}$, respectively.  
Thanks to the well-established calibrated relationships based on AGN single-epoch spectra (e.g., Kaspi et al.
2000, 2005; Wu et al. 2004; Peterson \& Bentz 2006; Marziani \& Sulentic 2012; Du et al. 2014, 2015; Peterson 2014; Wang et al.
2014), the black hole virial mass ($M_{\mathrm{BH}}$) and Eddington ratio ($L_{\mathrm{bol}}/L_{\mathrm{Edd}}$, where $L_{\mathrm{Edd}}=1.5\times10^{38}\,(M_{\mathrm{BH}}/{\it M}_\odot)\,\mathrm{erg\,s^{-1}}$ is the Eddington luminosity, Eq. 3.16 in Netzer (2013)) can be estimated in terms of the modeled H$\alpha$ broad emission line through the
 traditional method described by Wang et al. (2020a).    

Briefly, $M_{\mathrm{BH}}$ can be estimated by the calibration (Greene \& Ho 2007)
\begin{equation}
\small
  M_{\mathrm{BH}}=3.0\times10^6\bigg(\frac{L_{\mathrm{H\alpha}}}{10^{42}\ \mathrm{erg\ s^{-1}}}\bigg)^{0.45}\bigg(\frac{\mathrm{FWHM_{H\alpha}}}{1000\ \mathrm{km\ s^{-1}}}\bigg)^2\ {\rm M}_\odot
\end{equation}
and $L_{\mathrm{bol}}/L_{\mathrm{Edd}}$ through a bolometric correction of $L_{\mathrm{bol}}=9\lambda L_{\lambda}(5100\,{\rm \AA})$ (e.g., Kaspi et al. 2000), where (Greene \& Ho 2005)
\begin{equation}
\lambda L_\lambda(5100\,{\rm \AA})=2.4\times10^{43}\bigg(\frac{L_{\mathrm{H\alpha}}}{10^{42}\,\mathrm{erg\,s^{-1}}}\bigg)^{0.86}\ \mathrm{erg\,s^{-1}}\, .
\end{equation}

In deriving the intrinsic broad H$\alpha$ line luminosity $L_{\mathrm{H\alpha}}$ obtained at 
different epochs, the measured broad H$\alpha$ line fluxes of individual objects are first 
scaled by a factor determined by equaling the total [\ion{O}{3}] $\lambda$5007 line flux
to that of the SDSS DR16 spectrum, which is given by Wang (2015). The intrinsic extinction is then corrected from the 
narrow-line flux ratio $\mathrm{H\alpha/H\beta}$ by 
assuming the Balmer decrement of standard Case B recombination and a
Galactic extinction curve with $R_V=3.1$. 
Being dominated by the calibration scatter, the uncertainties of 
$M_{\mathrm{BH}}$ and $L_{\mathrm{bol}}/L_{\mathrm{Edd}}$ are $\sim 0.2$~dex and $\sim 65$\%
(or 0.28~dex), respectively. $L_{\mathrm{bol}}/L_{\mathrm{Edd}}$ is instead estimated from their unabsorbed X-ray 2--10~keV luminosity for the 
``turn-off'' state of two objects,  SDSS\,J124610.75+275615.9 and SDSS\,J151652.48+395413.4, by adopting 
a bolometric correction of $L_{\mathrm{bol}}=16L_{\mathrm{2-10~keV}}$.

\subsection{Stellar Population of the Hosts of the New CL-AGNs}

It is well known that both the 4000~\AA\ break [$D_{\rm n}(4000)$] and the equivalent width (EW) of the H$\delta$ absorption due to A-type stars ($\mathrm{H\delta_A}$) 
are widely used as reliable age indicators in AGN host galaxies until a few Gyr after a starburst  (e.g., Kauffmann et al. 2003; Heckman
et al. 2004; Kewley et al. 2006; Kauffmann \& Heckman 2009; Wild et al. 2010; Wang \& Wei 2008, 2010; Wang et al. 2013; Wang 2015), 
although both indices are sensitive to metallicity in very old stellar populations. 
The $D_{\rm n}(4000)$ is defined as (Balogh et al. 1999; Bruzual 1983)
\begin{equation}
 D_{\rm n}(4000)=\frac{\int_{4000}^{4100}f_\lambda d\lambda}{\int_{3850}^{3950}f_\lambda d\lambda}\, .
\end{equation}

The index H$\delta_{\mathrm A}$ is defined as (Worthey \& Ottaviani 1997)
\begin{equation}
   \mathrm{H\delta_A}=(4122.25–4083.50)\bigg(1-\frac{F_I}{F_c}\bigg)\ \mathrm{\AA}
\end{equation}
where $F_I$ is the flux within the feature bandpass of $\lambda\lambda$4083.50–4122.25,
and $F_c$ the flux of the pseudo continuum evaluated in the two beside regions: 
blue $\lambda\lambda$4041.60–4079.75 and red $\lambda\lambda$4128.50–4161.00.

Columns (11) ans (12) of Table 1 list the values of both $D_{\rm n}(4000)$ and H$\delta_{\mathrm A}$
that are measured from the modeled starlight component by Wang (2015).
Combining the uncertainties due to both measurements in duplicate observations (Wang et al. 2011)
and AGN continuum removal (Wang 2015), the typical uncertainties of $D_{\rm n}(4000)$ 
and H$\delta_{\mathrm A}$ are estimated to be $\sim 0.04$ and $\sim0.4$\AA, respectively.

\subsection{Statistics}
\subsubsection{SMBH Accretion}
Figure 4 shows the distributions of CL-AGNs in the $L_{\mathrm{bol}}$ vs. $M_{\mathrm{BH}}$ 
(left panels) and $L_{\mathrm{bol}}$
vs. $L_{\mathrm{bol}}/L_{\mathrm{Edd}}$ (right panels) diagrams, after combining the 9 
CL ``partially obscured'' AGNs identified in this study and the 
CL-AGNs studied and compiled by Wang et al. (2019, see references therein).  
The comparison samples 
shown by different symbols and colors are (1) the SDSS DR7 quasars with $z<0.5$ (Shen et al. 2011), (2) the SDSS DR3 narrow-line Seyfert 1 galaxies (NLS1s) 
given by Zhou et al. (2006), (3) the {\it Swift}/BAT AGN sample with a spectral type classification by Winter et al. (2012), and (4) the SDSS
intermediate-type Seyfert galaxies studied by Wang (2015). In the figure, the upper two panels correspond to the ``turn-on''
state, and the lower two panels to the ``turn-off'' state. 

Two facts can be gleaned from the figure. First, compared to the parent sample (i.e., the 170 
SDSS ``partially obscured'' AGNs studied by Wang (2015), the
magenta points in Figure 4),
one can see from the upper two panels that the ``partially obscured'' CL-AGNs identified by us tend 
to be biased against both high $L_{\mathrm{bol}}$ and high 
$L_{\mathrm{bol}}/L_{\mathrm{Edd}}$
as claimed previously for luminous quasars
(e.g., MacLeod et al. 2019; Wang et al. 2019; Frederick  et  al.  2019; Jin et al. 2022).
Second, SDSS\,J143016.03+230844.2 shows Balmer emission-line profiles that are typical of a 
Type~I AGN at  its ``turn-on'' state with high $L_{\mathrm{bol}}=8.6\times10^{45}\ \mathrm{erg\ s^{-1}}$ and $L_{\mathrm{bol}}/L_{\mathrm{Edd}}=0.1$, reinforcing our 
previous claim that 
there are two kinds of origins of intermediate-type AGNs: one to the well-accepted
orientation effect and the other to an intrinsic change of accretion rate (Wang et al. 2019).

\begin{figure*}[ht!]
\plotone{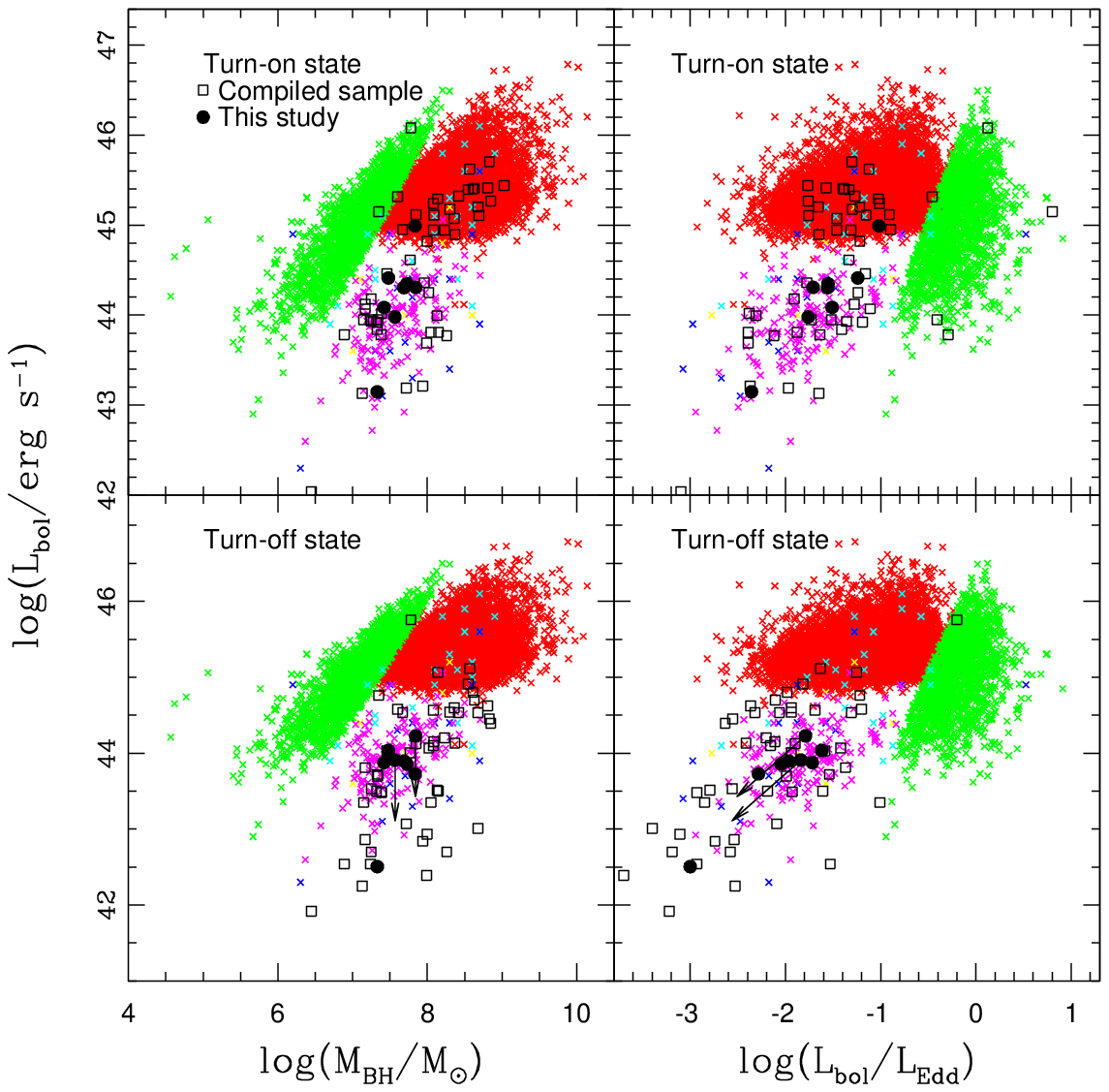}
\caption{Distributions of the CL-AGNs on the $L_{\mathrm{bol}}$--$M_{\mathrm{BH}}$ (left panels) and
$L_{\mathrm{bol}}$--$L_{\mathrm{bol}}/L_{\mathrm{Edd}}$ (right panels) diagrams, after combining the 9 new
``partially obscured'' CL-AGNs (denoted by the solid circles)
identified in this study and the 26 previously identified CL-AGNs (denoted by the open squares) complied by Wang et al. (2019). 
The ``turn-on'' and ``turn-off'' states are displayed in the upper and lower rows, respectively. 
The underlying comparison samples are described as follows. Red crosses, quasars with $z<0.5$ taken from the value-added SDSS DR7 quasar catalog 
(Shen et al. 2011); green crosses, the SDSS DR3 NLS1 catalog established by Zhou et al. (2006); and magenta crosses, the SDSS DR7 intermediate-type AGNs studied
by Wang (2015). The {\it Swift}/BAT AGN sample of Winter et al. (2012) is shown by the cyan, yellow, and blue crosses for Seyfert 1, 1.2, and 1.5 galaxies, respectively.
\label{fig:general}}
\end{figure*}

\subsubsection{Host Galaxies}

The $D_{\rm n}(4000)$ index measured by Wang (2015) ranges from 1.38 to 1.83 for 
the 9 ``partially obscured'' CL-AGNs. With a determined standard deviation of 0.15,
the average and median values of $D_{\rm n}(4000)$ are 1.58 and 1.59, respectively.
A comparison of the cumulative distribution of $D_{\rm n}(4000)$ is presented in Figure 5. 
A one-side Kolmogrov-Smirnov test returns a result that the distributions of the 9 CL-AGNs and 
of the parent sample (i.e., the 170 ``partially obscured'' AGNs studied by Wang (2015))
have a maximum distance of 0.40,
which corresponds to a discrepancy with a probability of 0.961.
The figure and the corresponding statistics 
suggest that the CL-AGNs identified in the current study are clearly
biased against a young stellar population, and 
tend to be associated with an intermediate-age stellar population,
although a larger sample of CL-AGNs with direct measurements of their host galaxies is needed to confirm these claims at a higher significance level.
A value of  $D_{\rm n}(4000) = 1.4$--1.6 is, in fact, usually adopted as a threshold for 
separating young and old stellar populations (e.g., Kauffmann et al. 2003).
The revealed association with an intermediate-age stellar population   
roughly agrees with the results recently reported by Liu et al. (2021) and Jin et al. (2022). 
Liu et al. (2021) point out that the
local CL-AGNs tend to reside in galaxies located in the ``green valley,'' rather than in blue host galaxies, where the average and median $D_{\rm n}(4000)$ values
are 1.47 and 1.46, respectively. The corresponding standard deviation of $D_{\rm n}(4000)$ is 0.21.  
In addition, based on stellar-population synthesis on 26 ``turn-off'' CL-AGNs, Jin et al. (2022) proposed that
CL-AGNs are mainly characterized by intermediate-age stellar populations. 

\begin{figure*}[ht!]
\plotone{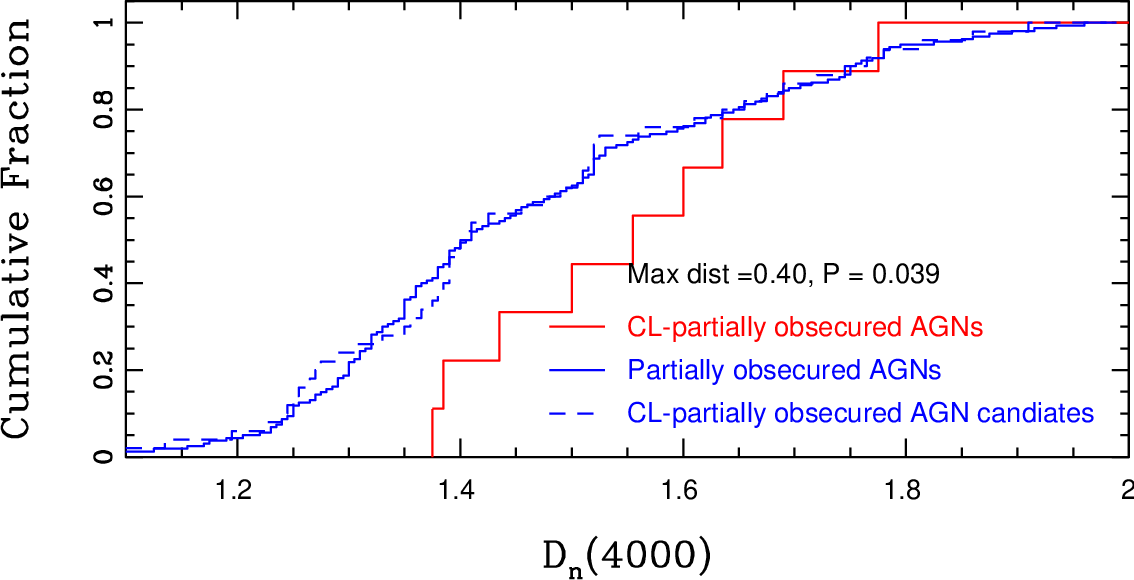}
\caption{A comparison of the distributions of $D_{\rm n}(4000)$ index. The distribution of the 9 
``partially obscured'' CL-AGNs is denoted by 
the red solid line. The blue dashed and blue solid lines show the distribution of the 59 ``partially obscured'' 
CL-AGN candidates that are selected according to their variability in the MIR band and the distribution of
the total 170 intermediate-type AGNs from which the candidates are selected.
\label{fig:general}}
\end{figure*}

As an additional test, Figure 6 shows the H$\delta_{\mathrm{A}}$ versus $D_{\rm n}(4000)$ plot, 
after combining the local CL-AGNs identified in this study and the ones quoted from literature. The values of both $D_{\rm n}(4000)$ and 
H$\delta_{\mathrm{A}}$ of 30 local CL-AGNs are reported in Table 3 in Liu et al. (2021).
For the 17 CL-AGNs recently identified in Lopez-Navas et al. (2023), the corresponding Lick indices
are obtained from the value-added MPA/JHU catalog (Kauffmann et al. 2003; Heckman \& Kauffmann 2006), 
in which the contamination caused by 
emission lines has been removed. One can see from the figure that the enlarged sample reinforces
the trend that CL-AGNs are associated with an intermediate-aged stellar population.

\begin{figure*}[ht!]
\plotone{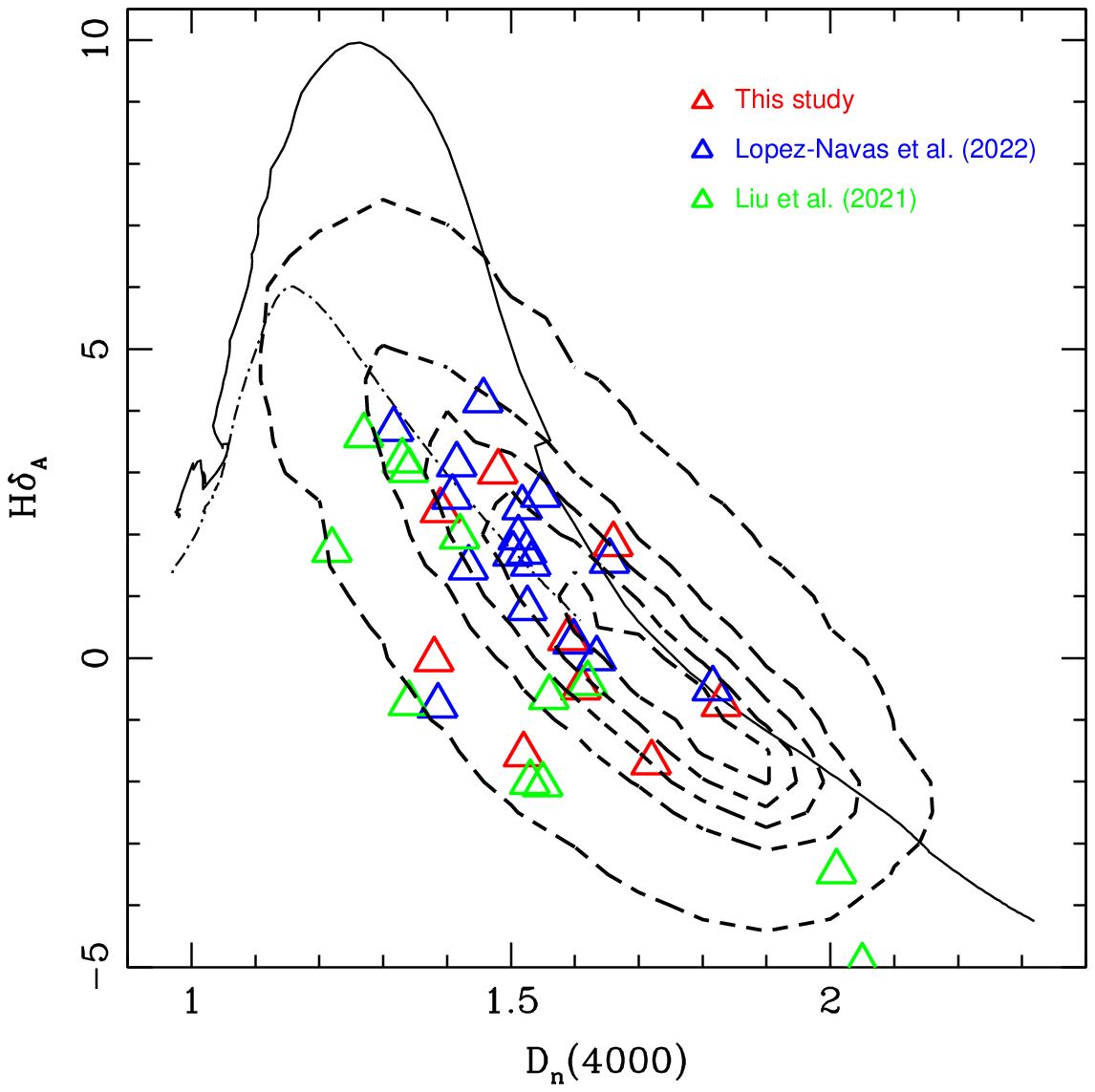}
\caption{$D_{\rm n}(4000)-\mathrm{H}\delta_{\mathrm{A}}$ plane for local CL-AGNs (see the main text 
for the details of the used samples). The underlying density contours (the dashed black heavy lines) 
show the distribution of $\sim80,000$ Seyfert 2 galaxies extracted from the 
MPA/JHU value-added catalog (e.g., Kauffmann et al. 2003; Heckman \& Kauffmann 2006). 
The solid line marks the stellar population evolution locus of the single-stellar population model 
with solar metallicity, and the dot-dashed line the model with exponentially decreasing star
formation rate $\psi(t)\propto e^{-(t/\mathrm{4Gyr})}$. 
\label{fig:general}}
\end{figure*}

\section{Discussion}

Through a follow-up spectroscopy program on the MIR variability-selected SDSS ``partially obscured'' AGNs,
we identified 9 CL-AGNs whose spectral types have changed during past $\sim 15$--20~yr from a parent sample of 59 candidates, which leads to a CL-AGN identification rate of 15\%, suggesting that 
variability in the MIR band is quite useful in searching for and studying the CL phenomenon. 
Taking into account the parent sample of 170 ``partially obscured'' AGNs, 
the CL-AGN fraction is determined to be no less
than 5\%. This number is much larger than the previously claimed CL-AGN fraction of 0.007--0.11\% 
(e.g., Yang et al. 2018; Yu et al. 2020). In addition, it is interesting that our spectroscopic campaign 
returns a ``turn-on'' over ``turn-off'' ratio of 5:4 that is quite close to 1. 
Miniutti et al. (2019) argued a connection between CL-AGNs and the rarely discovered X-ray
quasiperiodic eruptions (QPE) from SMBHs that predict similar numbers of ``turn-on'' and ``turn-off'' CL states
according to the symmetric light-curve profile of the eruptions, although the physical origin of the SMBH's QPE
is still an open question.

Subsequent spectral analysis of the  ``partially obscured'' CL-AGNs enables us to reinforce 
the previous claims that CL-AGNs tend to be biased against both high $L_{\mathrm{bol}}$ and high $L_{\mathrm{bol}}/L_{\mathrm{Edd}}$, 
and toward an intermediate-age stellar population. 

Figure 7 shows that the Keck follow-up spectrum with weak broad H$\alpha$ emission 
allows us to identify SDSS\,J151652.48+295413.4 as a new
repeat CL-AGNs with a rapid ``turn-on'' timescale of $\sim 2$\,yr. The repeat CL 
phenomenon is further supported by the X-ray emission level assessed from our new
{\it Swift}/XRT observation. The X-ray emission is 
close to that given by the {\it ROSAT} survey and leads to $L/L_{\mathrm{Edd}}=0.032$,
comparable to the value estimated in the ``turn-on'' state. 
Up to now, there have been only
eight confirmed repeat CL-AGNs: Mrk\,590, Mrk\,1018, NGC\,1566, NGC\,4151, NGC\,7603,
Fairall\,9, 3C\,390.3, and UGC\,3223 (Marin et al. 2019; Parker et al. 2019; Wang et al. 2020; Mathur et al. 2018). 
With $M_{\mathrm{BH}} = 3.7\times10^7\,{\rm M}_\odot$, the timescale of a typical TDE is predicted to be
$\Delta t=0.35(M_{\mathrm{BH}}/10^7\,{\rm M}_\odot)^{1/2}(M_\star/{\rm M}_\odot)^{-1}(R_\star/{\rm R}_\odot)^{3/2} \approx 1$~yr
(e.g., Rees 1988; Lodato \& Rossi 2011, and references therein) for the object, where $M_\star$ and $R_\star$ are respectively the mass and radius of the
disrupted star. Although this prediction is comparable to the ``turn-on'' timescale revealed in the object, it is difficult for the TDE scenario to explain the repeat
CL phenomenon identified in the object, because the TDE rate is expected to be as low as one event every $10^4$--$10^5$~yr per galaxy (e.g., Gezari et al. 2008; Donley et al. 2002; van Velzen et al. 2014).

\begin{figure*}[ht!]
\plotone{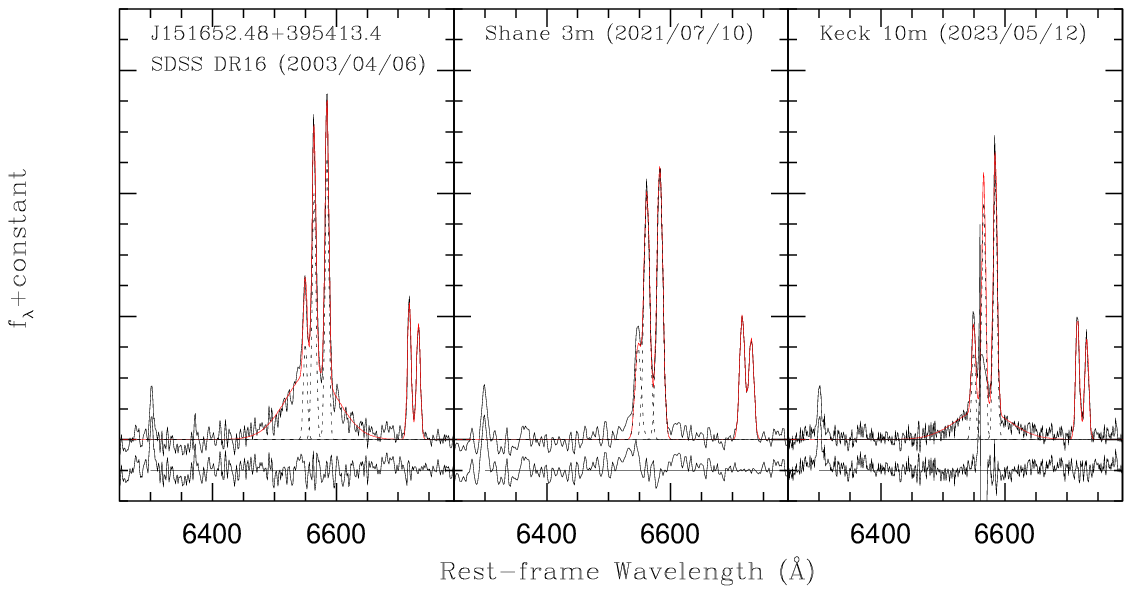}
\caption{A comparison of the line-profile modeling in the H$\alpha$ region of 
SDSS\,J151652.48+295413.4. The spectra obtained by SDSS, Shane 3~m, and Keck~II 10~m telescopes
at different epochs are displayed from left to right. In the modeling of the 
Keck spectrum, the narrow H$\alpha$ emission line is produced from the profile of 
the adjacent [\ion{N}{2}] $\lambda$6583 emission line by a fixed line flux ratio of 0.9
determined from the SDSS spectrum, because the narrow H$\alpha$ emission line falls in the 12~\AA\ chip gap that is ignored in our modeling. Symbols are the same as in Figure 2. 
\label{fig:general}}
\end{figure*}

\subsection{Physical Origin of the CL Phenomenon}

The MIR light curves of the ``partially obscured'' CL-AGNs identified in this study are shown in Figure 8, in which the $w1$ (3.4~$\mu$m) and $w2$ (4.6~$\mu$m) values are taken from the Wide-field Infrared Survey Explorer ({\it WISE} and {\it NEOWISE-R}; Wright et al. 2010; Mainzer et al. 2014)\footnote{We refer the 
readers to Wang et al. (2022) for details of SDSS\,J075244.20+455657.4 (B3\,0749+460A).}.
Table 3 compares the relative MIR brightness offsets $\delta w$ at the epoch of our spectroscopy 
with the identified CL status.  The relative offset is defined 
as $\delta w=(w-\langle w\rangle)/\Delta w$, where $\langle w\rangle$ is the mean brightness and 
$\Delta w$ the photometric error, respectively. A positive $\delta w$ denotes a MIR brightness fainter
than the mean brightness, and a negative one a brighter MIR brightness.
Although the epochs of the SDSS DR16 spectra are not covered by the {\it WISE}\ survey, one can see 
from the table that the CL status revealed by 
our follow-up spectra is generally related to the $w1$ and $w2$ brightness: a ``turn-on'' status corresponds to an MIR brightening, and a `turn-off'' status to an MIR dimming, except
SDSS\,J102530.44+462808.6 and SDSS\,J124610.75+275615.9. A close inspection of the light curves
of both objects shows that their CL status are related with the light curve trends at the corresponding
spectroscopic epochs: the ``turn-on'' phenomenon occurs during the MIR brightening process, and 
the ``turn-off'' phenomenon during the MIR dimming process.

\begin{table}
        \centering
        \caption{MIR brightness versus CL status}
        \label{tab:example_table}
        \begin{tabular}{ccccc} 
        \hline
        \hline
        Object & MJD &  $\delta w1$  & $\delta w2$ & CL status \\
               & (day) &            &      & \\
          (1) & (2) & (3) & (4) & (5) \\     
        \hline      
        J075244.20+455657.4 & 57428 & +1.8 & +3.9 & off \\
        J102530.44+462808.6 & 59345 & +0.9 & +2.0 & on  \\
        J124610.75+275615.9 & 59407 & $-0.1$ & $-0.4$ & off \\
        J141020.59+130829.3 & 59359 & $-0.8$ & $-1.2$ & on  \\
        J143016.03+230344.2 & 59345 & $-6.3$ & $-7.2$ & on  \\
        J145536.95+013151.2 & 59407 & $-2.5$ & $-2.1$ & on \\
        J150240.63+612851.4 & 59407 & $-0.9$ & $-0.6$ & on \\
        J151652.48+395413.4 & 59405 & +1.0 & +2.3 & off \\
        J161950.25+261329.6 & 59724 & +0.8 & +1.5 & off \\
        \hline
        \end{tabular}
        \tablecomments{Columns (1): Object SDSS ID; Column (2): Modified Julian date when our spectroscopy
        was carried out; Columns (3) and (4): Relative MIR brightness offsets in \it WISE\rm\ $w1$ and 
        $w2$ bands at the epoch listed in Column (2), see the main text for the definition; 
        Column (5): CL status reported in Table 1.}
\end{table}

The dependence of CL status on MIR brightness is \rm consistent with the expectation of the accretion-rate enhancement scenario of the CL phenomenon (e.g.,
Sheng et al. 2017; Stern et al. 2018; Wang et al. 2019, 2020, 2021; Yang et al. 2018; Lopez-Navas et al. 2022, 2023).
In addition to their MIR variation, the spectropolarimetry performed by Hutsemekers et al. (2019) 
shows that CL-AGNs in the ``turn-off'' state typically have polarization below 1\%, implying that the clumpy 
obscuration scenario is unlikely to explain the disappearance of the broad emission lines.

The fact that CL-AGNs tend to be biased against high $L_{\mathrm{bol}}$ and high $L_{\mathrm{bol}}/L_{\mathrm{Edd}}$ therefore motivates 
some authors to argue that the CL phenomenon could be understood by the disk-wind broad-line region (BLR) models previously proposed by
Elitzur \& Ho (2009) and Nicastro (2000).  On the one hand, in the Elitzur \& Ho (2009) model, because the mass-outflow rate scales with $L$ as $L^{1/4}$, 
an observable BLR cannot be sustained below a certain luminosity, $L\approx5\times10^{39}(M_{\mathrm{BH}}/10^7~{\rm M}_\odot)^{2/3}\,\mathrm{erg\,s^{-1}}$,
which corresponds to a critical value of $L_{\mathrm{bol}}/L_{\mathrm{Edd}} \approx 10^{-6}$ (Elitzur \& Shlosman 2006).

On the other hand, a much higher critical value of $L_{\mathrm{bol}}/L_{\mathrm{Edd}} \approx 2$--3 $\times 10^{-3}$ has been proposed 
for $M_{\mathrm{BH}}$ within a range of $10^{7-8}~{\rm M}_\odot$ by Nicastro (2000), in which the appearance or
disappearance of the BLR depends on a critical radius of the accretion disk where the power deposited into the vertical outflow is the maximum value.
One would argue against this scenario by being aware of the fact that, as shown in Figure 3
and Table 1, the calculated $L_{\mathrm{bol}}/L_{\mathrm{Edd}}$ at the 
``turn-off'' state is usually higher than the above critical value by an order of magnitude. This problem can be easily resolved since the reported
$L_{\mathrm{bol}}/L_{\mathrm{Edd}}$ at the ``turn-off'' state is usually estimated from the weakened broad Balmer emission lines. In fact, the X-ray inferred $L_{\mathrm{bol}}/L_{\mathrm{Edd}}$ is estimated to be $\sim2\times10^{-3}$ 
and $\sim1\times10^{-3}$ for SDSS\,J075244.20+455657.4 (Wang et al. 2022) 
and SDSS\,J151652.48+395413.4 (this study), respectively, at 
their ``turn-off'' state with a disappeared classic BLR. In addition,
our X-ray observation of the CL-AGN UGC\,3223 shows that its $L_{\mathrm{bol}}/L_{\mathrm{Edd}}$ was as low as $\sim 2 \times 10^{-4}$ when 
the broad Balmer emission lines disappeared completely (Wang et al. 2020b, and 
see  Ai et al. (2020) for other cases 
with $L_{\mathrm{bol}}$ and $L_{\mathrm{bol}}/L_{\mathrm{Edd}}$ inferred from X-rays).

In the disk-wind model, a thermally unstable radiation-pressure-dominated disk region is required for
the existence of a BLR, which implies that the thermal timescale is typical of the CL phenomenon.
This timescale is, in fact, able to account for the observed variations of order 10--20~yr, especially the $\sim 2$~yr rapid ``turn-on'' process observed in 
SDSS\,J151652.48+395413.4. 
The evolutionary $\alpha$-disk model developed by Siemiginowska et al. (1996), in fact, predicts 
\begin{equation}
\small
 t_{\mathrm{th}} \approx \frac{1}{\alpha\Omega_{\mathrm{K}}}\\
 =2.7\bigg(\frac{\alpha}{0.1}\bigg)^{-1}\bigg(\frac{r}{10^{16}\,\mathrm{cm}}\bigg)^{3/2}\bigg(\frac{M_{\mathrm{BH}}}{10^8~{\rm M}_\odot}\bigg)^{-1/2} \mathrm{yr}, 
\end{equation}
which leads to a variability timescale of 3--8~yr for the 9 ``partially obscured'' AGNs, when the fiducial values of
``viscosity parameter'' $\alpha=0.1$ and $r=10^{16}~\mathrm{cm}$ and the 
estimated $M_{\mathrm{BH}}$ are adopted. Additionally, 
a shorter variability timescale can be obtained either by 
introducing a narrow unstable zone (Sniegowska et al. 2020),   
or by involving an accretion disk elevated by a magnetic field (e.g., Ross et al. 2018; Stern et al. 2018; Dexter \& Begelman 2019), which is supported by recent numerical simulations carried out by Pan et al. (2021).
 
On the contrary, the classical viscous radial inflow is 
not a plausible scenario for the observed CL phenomenon.
The viscous timescale of a viscous radial inflow can be estimated as (e.g., Shakura \& Sunyaev 1973; Krolik 1999; LaMassa et al. 2015; Gezari et al. 2017)
\begin{equation}
\scriptsize
 t_{\mathrm{infl}} = 6.5\bigg(\frac{\alpha}{0.1}\bigg)^{-1}\bigg(\frac{L/L_{\mathrm{Edd}}}{0.1}\bigg)^{-2}\bigg(\frac{\eta}{0.1}\bigg)^2\bigg(\frac{r}{r_g}\bigg)^{7/2}\bigg(\frac{M_{\mathrm{BH}}}{10^8~{\rm M}_\odot}\bigg)~\mathrm{yr}\, ,
\end{equation}
where $\eta$ is the efficiency of converting potential energy to radiation and $r_g$ is the
gravitational radius in units of $GM/c^2$. Taking the fiducial values of $\alpha=\eta=0.1$, the 
viscous timescale is estimated to be $10^{3-5}$~yr, when $r\approx(50$--100)$r_g$ is 
adopted to account for the outer disk producing optical emission. However,
Feng et al. (2021b) propose that the inflow timescale can be significantly reduced
in a magnetic accretion disk-outflow model by magnetic outflows.

With increasing cases of  identified CL-AGNs,  a few other possible models have been proposed recently to understand the CL phenomenon. For instance,
the theoretical study by Wang \& Bon (2020) suggests that CL-AGNs could be triggered by 
close binaries of SMBHs with a high eccentricity. A tidal torque on the mini-disk of each SMBH can either 
squeeze or expand the disk, resulting in a spectral type transition cycle determined by the orbital period.    
The authors argue that this scenario is able to explain the highly asymmetric and double-peaked broad-line
profiles observed in some CL-AGNs (e.g., Storchi-Bergmann et al. 2017).

\begin{figure*}[ht!]
\plotone{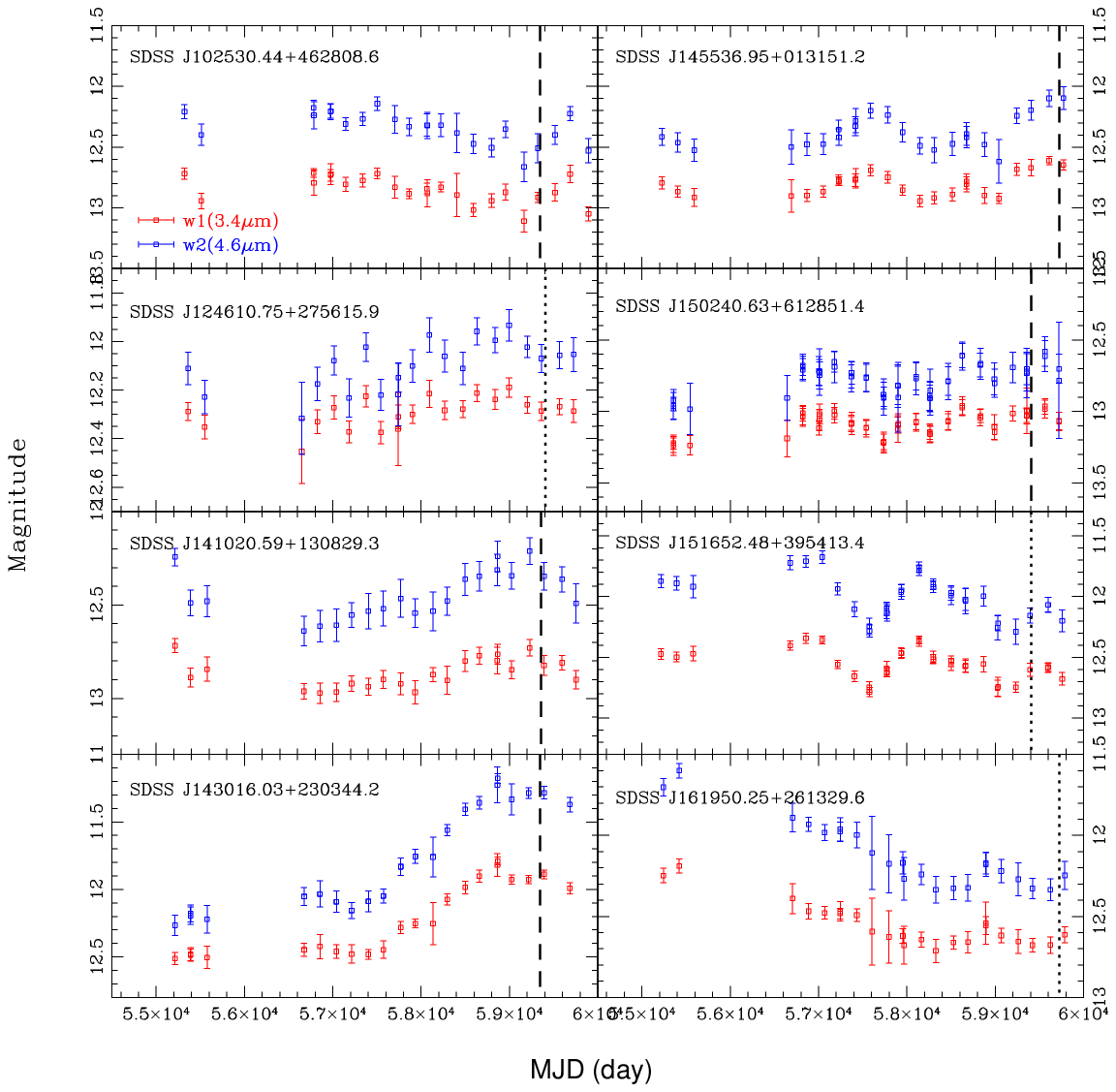}
\caption{{\it WISE} MIR light curves of the 8 ``partially obscured'' CL-AGNs. Each light curve is binned by averaging the measurements within one day.
In each panel, the vertical black lines mark the epochs of our optical spectra, where the long-dashed line corresponds to a ``turn-on'' state and the short-dashed 
one to a ``turn-off'' state.
\label{fig:general}}
\end{figure*}

\subsection{Implication from the Intermediate-Age Stellar Populations}

Although no significant difference between the host galaxies of CL-AGNs and NCL-AGNs has been reported by some studies in recent years
(e.g., Charlton  et  al. 2019; Yu et al. 2020; Dodd et al. 2021), 
our spectroscopic campaign on ``partially obscured'' CL-AGNs and direct measurements
of stellar populations allow us to suggest 
that CL-AGNs tend to be dominated by intermediate-age populations, roughly consistent with Liu et al. (2021) and Jin et al. (2022).

Given the inferred intermediate-age stellar population and the concept of coevolution between SMBH and host galaxy (see Heckman \& Kauffmann 2011, Heckman \& Best 2014 for a review), we propose that CL-AGNs are AGNs at a specific evolutionary stage. Figure 9 marks the 9 ``partially obscured'' CL-AGNs, 
along with other CL-AGNs identified and complied in previous studies (Lopez-Navas et al. 2022; Liu et al. 2021 and references therein), \rm 
on the $D_{\rm n}(4000)$--$L_{\mathrm{bol}}/L_{\mathrm{Edd}}$ diagram.
There were, in fact, a series of studies \rm that show 
a strong relationship between $L_{\mathrm{bol}}/L_{\mathrm{Edd}}$ and $D_{\rm n}(4000)$ for
local AGNs (Wang 2015 and references therein). 
The relationship suggests a coevolution between SMBH growth and host galaxy wherein the SMBH resides:  
$L_{\mathrm{bol}}/L_{\mathrm{Edd}}$ decreases as the young stellar population continuously ages (e.g., Kewley et al. 2006; Wang et al. 2006, 2013; Wang \& Wei 2008, 2010; Wang 2015).
It is clearly that the local CL-AGNs closely follow the $D_{\rm n}(4000)$--$L_{\mathrm{bol}}/L_{\mathrm{Edd}}$ sequence, and tend to mainly occupy the middle of the sequence simply 
due to their associated intermediate-aged stellar populations, \rm which implies that CL is a special phenomenon occurring only in 
evolved AGNs in the context of the coevolution of SMBHs and their host galaxies.

By analyzing the distribution of $L_{\mathrm{bol}}/L_{\mathrm{Edd}}$ as a function of $M_{\mathrm{BH}}$ and 
properties of the host galaxies for a large sample 
of local Seyfert 2 galaxies surveyed by SDSS,  Kauffmann \& Heckman (2009) proposed that there are two distinct regimes of SMBH accretion. 
On the one hand, the growth of SMBHs associated with a young central stellar population and significant star formation is less dependent on
the central stellar population of the galaxy. The plentiful cold gas supply can guarantee a steady inward fueling flow of cold gas (i.e., ``feast fueling''). 
One the other hand,   
for the SMBHs associated with an old central stellar population, the time-averaged mass growth rate is proportional to the mass of the bulge of the host galaxy
(i.e., ``famine fueling'').  
This dependence could be understood if, when the cold gas in a reservoir is consumed, 
the SMBHs are fed by slow stellar winds generated by evolved stars or by inward mass transport 
triggered by either minor mergers or multiple collisions of cold-gas clumps in the intergalactic medium 
(e.g., Kauffmann \& Heckman 2009; Davies et al. 2007, 2014; Pizzolato \& Soker 2005; Gaspari et al. 2013).

\begin{figure}[ht!]
\plotone{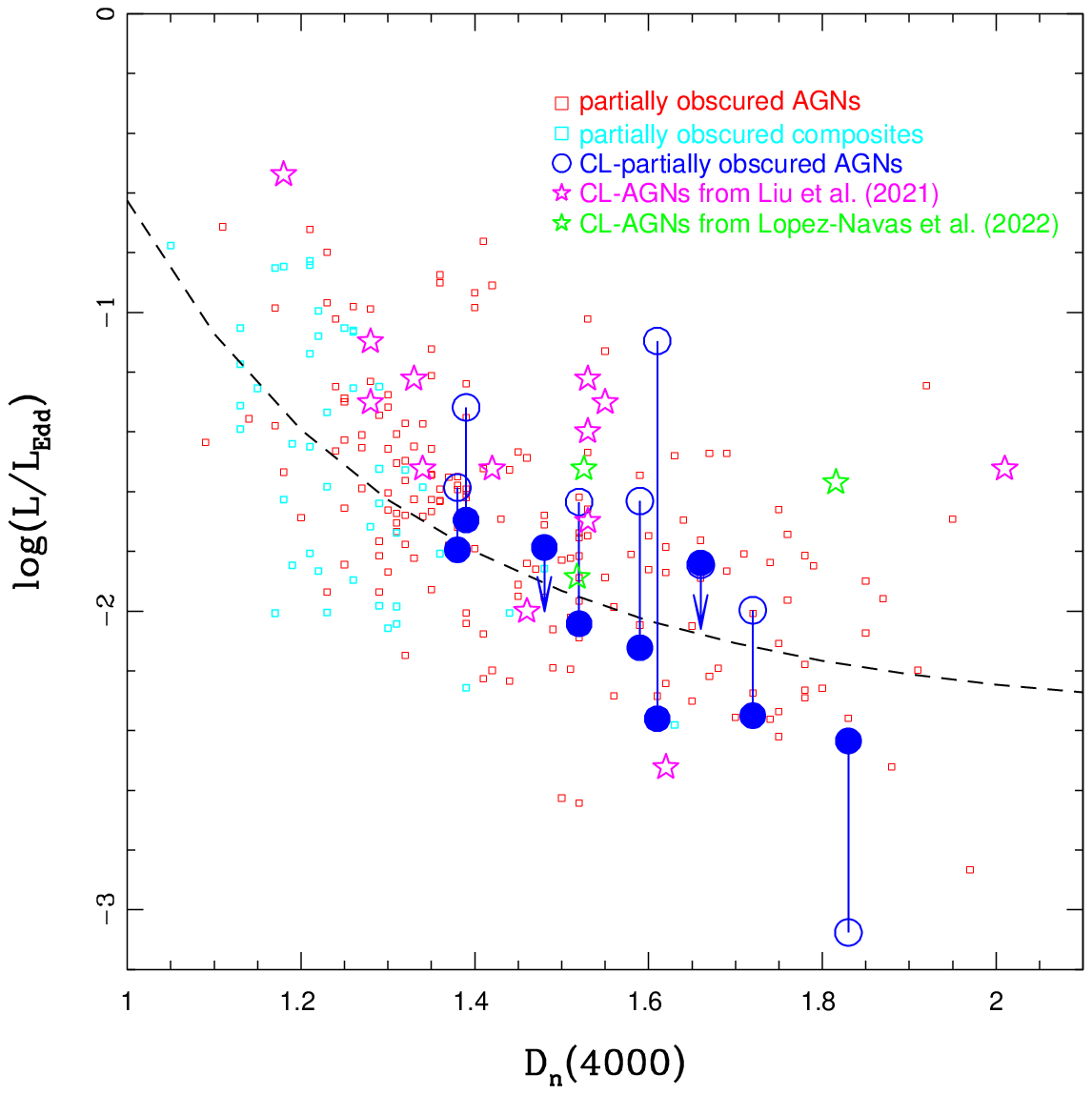}
\caption{$L_{\mathrm{bol}}/L_{\mathrm{Edd}}$ plotted as a function of stellar population age index $D_{\rm n}(4000)$. The 9 ``partially obscured'' CL-AGNs identified by us are shown 
by the solid and open blue circles for the ``turn-off'' and ``turn-on'' states, respectively. The other CL-AGNs (at ``turn-on'' state) 
compiled in Liu et al. (2021 and references therein) 
and identified in Lopez-Navas et al. (2022) are displayed by the open stars in magenta and green, respectively.
The small red and cyan open squares are the local ``partially obscured'' Seyfert galaxies and composite galaxies studied by Wang (2015), respectively.  
The dashed line is the best-fit nonlinear relationship to these ``partially obscured'' AGNs. 
\label{fig:general}}
\end{figure}

Given the inferred stellar population and the concept of coevolution between SMBH and host galaxy, 
here we propose that CL-AGNs are probably at a transition between the feast and famine SMBH fueling stages that are profiled by Kauffmann \& Heckman (2009). The authors, in fact, indicate that a transition between the aforementioned two SMBH accretion regimes occurs at 
$D_{\rm n}(4000)=1.5$--1.8, comparable to the measured average value and standard deviation of $D_{\rm n}(4000)$ of the CL-AGNs focused in the current study.
At the transition, the cold gas in the reservoir has been almost exhausted by steady inward fueling flow;
thereafter, episodic fueling is expected when the fueling is contributed by 
slow stellar winds of evolved stars, especially those on the asymptotic giant branch, or by chaotic accretion (e.g., King \& Pringle 2006, 2007). 
The proposed particular evolutionary stage of CL-AGNs is supported by recent other 
studies. On the one hand, from the view of host galaxies, besides the studies of direct stellar populations
mentioned above, Dodd et al. (2021) indicate that CL-AGNs tend to reside in high-density pseudobulges 
located in the so-called ``green valley'' that is believed to be a transition region between
active star-forming galaxies and inactive elliptical galaxies owing to the quenching of AGNs. On
the other hand, from the view of SMBH accretion, the MIR Eddington ratio of CL-AGNs
is found to be in accord with a transition between a Shakura-Sunyaev disk (Shakura \& Sunyaev
1973) and radiatively inefficient accretion flow (Lyu et al. 2022). By comparing CL-AGNs and
different types of AGNs, Liu et al. (2021) suggest that CL-AGNs are possible at a specific evolutionary stage with $L_{\mathrm{bol}}/L_{\mathrm{Edd}}\approx0.1$, since the objects show a transition between the 
positive and negative $\Gamma–L_{\mathrm{X}}/L_{\mathrm{Edd}}$ correlation in the CL phenomenon.

\section{Conclusions}

We study the physics of CL-AGNs from the perspective of the coevolution of AGNs and their host galaxies.
By focusing on the local ``partially obscured'' AGNs, our follow-up spectroscopy and 
detailed analysis of these spectra allow us to arrive at  following conclusions.
\begin{enumerate}
\item Nine new ``partially obscured'' CL-AGNs are identified from a sample of 59 candidates that are selected according 
to their variability in the MIR band.
\item The host galaxies of the identified ``partially obscured'' CL-AGNs are biased against 
 young stellar populations, and tend to be dominated by intermediate-age stellar populations, consistent with the results recently reported by
Liu et al. (2021) and Jin et al. (2022). This population motivates us to propose that CL-AGNs are probably AGNs at a specific evolutionary stage,
going from a plentiful supply of cold gas to fueling by the slow stellar winds of evolved stars --- that is, at a transition stage from ``feast'' to ``famine'' SMBH fueling.
The validation of the evolutionary-stage specificity of CL-AGNs requires larger samples of CL-AGNs with directly measured host-galaxy properties. Future studies using such samples would illuminate the relationship between CL-AGN behavior and host-galaxy properties, allowing us to gain deeper insights into the unique evolutionary mechanisms of these objects.
\item We reinforce the claims that CL-AGNs tend to be biased against a high 
$L_{\mathrm{bol}}/L_{\mathrm{Edd}}$ and a high $L_{\mathrm{bol}}$, 
implying that the disk-wind BLR model is a plausible explanation of the CL phenomenon.

\end{enumerate}

\acknowledgments

The authors would like to thank the anonymous referees for his/her careful review and helpful comments
that improve our study significantly.
This study is supported by the National Natural Science Foundation of China under grants 12173009 and 12273054, the Natural Science Foundation of Guangxi (2020GXNSFDA238018), 
and the Strategic Pioneer Program on Space Science, Chinese Academy of Sciences, grants XDA15052600 and XDA15016500.
The authors are grateful for support from the National Key Research and Development Project of China (grant 2020YFE0202100). 
A.V.F.'s group at U.C. Berkeley has received financial assistance from the Christopher R. Redlich Fund, Alan Eustace (W.Z. is a Eustace Specialist in Astronomy), Briggs and Kathleen Wood (T.G.B. is a Wood Specialist in Astronomy), and many other donors.

We acknowledge the support of the staff of the  Xinglong
2.16m telescope. This work was partially supported by the  Open  Project  Program  of  the  Key  Laboratory  of  Optical  Astronomy,  National Astronomical Observatories, Chinese Academy of Sciences.
Some of the data presented herein were obtained at the W. M. Keck Observatory, which is operated as a scientific partnership among the California Institute of Technology, the University of California, and the National Aeronautics and Space Administration (NASA); the observatory was made possible by the generous financial support of the W. M. Keck Foundation. 
The Kast red CCD detector upgrade on the Shane 3~m telescope at Lick Observatory, led by B. Holden, was made possible by the Heising–Simons Foundation, William and Marina Kast, and the University of California Observatories. Research at Lick Observatory is partially supported by a generous gift from Google.    
 
We thank the {\it Swift} Acting PI, Brad Cenko,
for approving our target-of-opportunity request, as well as the
{\it Swift} observation team for assistance.
This study uses the SDSS data archive created and distributed by the
Alfred P. Sloan Foundation, the Participating Institutions, the U.S.
National Science Foundation, and the U.S. Department of
Energy Office of Science.
This study used the NASA/IPAC Extragalactic Database (NED), which is operated by the Jet Propulsion
Laboratory, California Institute of Technology. 
It also used data collected by the {\it Wide-field Infrared Survey Explorer (WISE)}, which is a joint project
of the University of California at Los Angeles and the Jet Propulsion Laboratory/California Institute of
Technology, funded by NASA.

\vspace{5mm}
\facilities{Lick Shane 3~m telescope (Kast), NAOC Xinglong 2.16~m telescope, Keck~II 10~m 
telescope (DEIMOS), {\it Swift}/XRT}
\software{IRAF (Tody 1986, 1992), IDL, HEASOFT, XSPEC}
%

%
%
%


\clearpage
\begin{thebibliography}{}


\bibitem[Ai et al. (2020)]{Ai20} Ai, Y. L., Dou, L. M., Yang, C. W., et al. 2020, \apjl, 890, 29  
\bibitem[Antonucci (1993)]{ant93} Antonucci, R. R. J. 1993, \araa, 31, 473
\bibitem[Arnaud (1996)]{arn96} Arnaud, K. A. 1996, ASPC, 101, 17
\bibitem[Balogh et al. (1999)]{bal99} Balogh, M.L. et al. 1999, \apj, 527, 54
\bibitem[Bentz et al. (2013)]{bez13} Bentz, M. C., Denney, K. D., Grier, C. J., et al. 2013, \apj, 767, 149
\bibitem[Blanchard et al. (2017)]{bla17} Blanchard, P. K., Nicholl, M., Berger, E., et al. 2017, \apj, 843, 106
\bibitem[Boroson (2005)]{bor05} Boroson, T. A. 2005, \aj, 130, 381
\bibitem[Bowen \& Vaughan (1973)]{bov73} Bowen, I. S., \& Vaughan, A. H., Jr. 1973, ApOpt, 12, 1430
\bibitem[Bruhweiler \& Verner (2008)]{brv08} Bruhweiler, F., \& Verner, E. 2008, \apj, 675, 83
\bibitem[Bruzual (1983)]{bru83} Bruzual, A.G. 1983, \apj, 273, 105
\bibitem[Bruzual \& Charlot (2003)]{brc03} Bruzual, G., \& Charlot, S. 2003, \mnras, 344, 1000
\bibitem[Cardelli et al. (1989)]{car89} Cardelli, J. A., Clayton, G. C., \& Mathis, J. S. 1989, \apj, 345, 245
\bibitem[Charlton et al. (2019)]{cha19} Charlton, P. J. L., Ruan, J. J., Haggard, D., et al. 2019, \apj, 876, 75
\bibitem[Cooper et al. (2012)]{coo12} Cooper, M. C., Newman, J. A., Davis, M., Finkbeiner, D. P., \& Gerke, B. F. 2012, spec2d, ASCL, 1203.003
\bibitem[Davies et al. (2017)]{dav17} Davies, R. I., Hicks, E. K. S., Erwin, P., et al. 2017, \mnras, 466, 4917
\bibitem[Davies et al. (2014)]{dav14} Davies, R. I., Maciejewski, W., Hicks, E. K. S., et al. 2014, \apj, 792, 101
\bibitem[Dexter \& Begelamn (2019)]{deb19} Dexter, J., \&  Begelman, M. C. 2019, \mnras, 483, L17  
\bibitem[Dietrich et al. (2002)]{die02} Dietrich, M., Hamann, F., Shields, J. C., et al. 2002, \apj, 581, 912
\bibitem[Dimitrijevic et al. (2007)]{dim07} Dimitrijevic, M. S., Popovic, L. C., Kovacevic, J., Dacic, M., \& Ilic, D. 2007, \mnras, 374, 1181
\bibitem[Dodd et al. (2021)]{dod21} Dodd, S. A., Law-Smith, J. A. P., Auchettl, K., et al. 2021, \apjl, 907, 21
\bibitem[Donley et al. (2002)]{don02} Donley, J. L., Brandt, W. N., Eracleous, M., \& Boller, T. 2002, \aj, 124, 1308 
\bibitem[Du et al. (2015)]{dup15} Du, P., Hu, C., Lu, K. X., et al. 2015, \apj, 806, 22
\bibitem[Du et al. (2014)]{dup14} Du, P., Wang, J. M., Hu, C., et al. 2014, \mnras, 438, 2828
\bibitem[Elitzur (2012)]{eli12} Elitzur, M., 2012, \apjl, 747, 33 
\bibitem[Elitzur \& Ho (2009)]{elh09} Elitzur, M., \& Ho, L. C. 2009, \apjl, 701, 91
\bibitem[Elitzur et al. (2014)]{eli14}  Elitzur, M., Ho, L. C., \& Trump, J. R. 2014, \mnras, 438, 3340  
\bibitem[Elitzur \& Shlosman (2006)]{els06} Elitzur, M., \& Shlosman, I. 2006,  \apjl, 648, 101
\bibitem[Faber et al. (2003)]{fab03} Faber, S. M., Phillips, A. C., Kibrick, R. I., et al. 2003, SPIE, 4841, 1657
\bibitem[Feng et al. (2021a)]{fen21a} Feng, H. C., Hu, C., Li, S. S., et al. 2021a, \apj, 909, 18
\bibitem[Feng et al. (2021b)]{fen21b} Feng, J., Cao, X., Li, J. W., \& Gu, W. M. 2021b, \apj, 916, 61 
\bibitem[Filippenko (1982)]{fil82} Filippenko, A. V. 1982, \pasp, 94, 715
\bibitem[Francis et al. (1992)]{fra92} Francis, P. J., Hewett, P. C., Foltz, C. B., \& Chaffee, F.H. 1992, \apj, 398, 476
\bibitem[Frederick et al. (2019)]{fre19} Frederick, S., Gezari, S., Graham, M. J., et al. 2019, \apj, 883, 31 
\bibitem[Gaspari et al. (2013)]{gas13} Gaspari, M., Ruszkowski, M., \& Oh, S. P. 2013, \mnras, 432, 3401 
\bibitem[Gezari et al. (2008)]{gez08} Gezari, S., Basa, S., Martin, D. C., et al. 2008, \apj, 676, 944 
\bibitem[Gehrels et al. (2004)]{geh04} Gehrels, N., Chincarini, G., Giommi, P., et al. 2004, \apj, 611, 1005
\bibitem[Gezari et al. (2017)]{gez17} Gezari, S., Hung, T., Cenko, S. B., et al. 2017, \apj, 835, 144
\bibitem[Guo et al. (2019)]{guo19} Guo, H. X., Sun, M. Y., Liu, X., Wang, T. G., Kong, M. Z., Wang, S., Sheng, Z. F., \& He, Z. C. 2019, \apjl, 833, 44 
\bibitem[Graham et al. (2020)]{gra20} Graham, M. J., Ross, N. P., Stern, D., et al. 2020, \mnras, 491, 4925
\bibitem[Green et al. (2022)]{gre22} Green, P. J., Pulgarin-Duque, L., Anderson, S. F., et al. 2022, arXiv: astro-ph/220109123
\bibitem[Greene \& Ho (2005)]{grh05} Greene, J. E., \& Ho, L. C. 2005, \apj, 630 ,122
\bibitem[Greene \& Ho (2007)]{grh07} Greene, J. E., \& Ho, L. C. 2007, \apj, 670, 92
\bibitem[Grimm et al. (2003)]{gri03} Grimm, H. -J., Gilfanov, M., \& Sunyaev, R. 2003, \mnras, 339, 793
\bibitem[Gunn et al. (2006)]{gun06} Gunn, J. E., Siegmund, W. A., Mannery, E. J., et al. 2006, \aj, 131, 2332
\bibitem[Haoc et al. (2005)]{hao05} Hao, L., et al. 2005, \aj, 129, 1795
\bibitem[Harrison et al. (2014)]{har14} Harrison, C. M., Alexander, D. M., Mullaney, J. R., \& Swinbank, A. M. 2014, \mnras, 441, 3306
\bibitem[Heckman (1980)]{heckman80} Heckman, T. M. 1980, \aap, 87, 152
\bibitem[Heckman \& Best (2014)]{keb14} Heckman, T. M., \& Best, P. N. 2014, \araa, 52, 589
\bibitem[Heckman \& Kauffmann (2006)]{hek06} Heckman, T. M., \& Kauffmann, G. 2006, \nar, 50, 677
\bibitem[Heckman \& Kauffmann (2011)]{hek11} Heckman, T. M., \& Kauffmann, G. 2011, Science, 333, 182
\bibitem[Heckman et al. (2004)]{hec04} Heckman, T. M., Kauffmann, G., Brinchmann, J., Charlot, S., Tremonti, C., \& White, S. D. M. 2004, \apj, 613, 109
\bibitem[Hon et al. (2020)]{hon20} Hon, W. J., Webster, R., \& Wolf, C. 2020, \mnras, 497, 192
\bibitem[Hon et al. (2022)]{hon22} Hon, W. J., Wolf, C.,  Onken, C. A., Webster, R., \& Auchettl, K. 2022, \mnras, 511, 54 
\bibitem[Husemann et al. (2016)]{hus16} Husemann, B., Urrutia, T., Tremblay, G. R., et al. 2016, \aap, 593, L9
\bibitem[Hutsemekers et al. (2019)]{hut19} Hutsemekers D., Agis Gonzalez B., Marin F., Sluse D., Ramos Almeida C., Acosta Pulido J.-A., 2019, \aap, 625, A54
\bibitem[Jarvela et al. (2017)]{jar17} Jarvela, E., Lahteenmaki, A., Lietzen, H., Poudel, A., Heinamaki, P., \& Einasto, M. 2017, \aap, 606, 9
\bibitem[Jin et al. (2022)]{jin22} Jin, J. J., Wu, X. B., \& Feng, X. T. 2022, \apj, 926, 184  
\bibitem[Kalberla et al. (2005)]{kal05} Kalberla, P. M. W., Burton, W. B., Hartmann, D., et al. 2005, \aap, 440, 775
\bibitem[Kaspi et al. (2005)]{kas05} Kaspi, S., Maoz, D., Netzer, H., et al. 2005, \apj, 629, 61
\bibitem[Kaspi et al. (2000)]{kas00} Kaspi, S., Smith, P. S., Netzer, H., et al. 2000, \apj, 533, 631
\bibitem[Kauffmann \& Heckman (2009)]{kah09} Kauffmann, G., \& Heckman, T. M. 2009, \mnras, 397, 135
\bibitem[Kauffmann et al. (2003)]{kau03} Kauffmann, G., Heckman, T. M., White, S. D. M., et al. 2003, \mnras, 341, 33
\bibitem[Kewley et al. (2006)]{kew06} Kewley, L. J., Groves, B., Kauffmann, G., \& Heckman, T. M. 2006, \mnras, 372, 961
\bibitem[King \& Pringle (2006)]{kip06} King, A. R., \& Pringle, J. E. 2006, \mnras, 373, L90
\bibitem[King \& Pringle (2007)]{kip07} King, A. R., \& Pringle, J. E. 2007, \mnras, 377, L25
\bibitem[Kollatschny et al. (2020)]{kol20} Kollatschny, W., Grupe, D., Parker, M. L., et al. 2020, \aap, 638, 91
\bibitem[Kollatschny et al. (2018)]{kol18} Kollatschny, W., Ochmann, M. W., Zetzl, M., Haas, M., Chelouche, D., Kaspi, S., Pozo Nuñez, F., \& Grupe, D. 2018, \aap, 619, 168 
\bibitem[Kollmeier et al. (2017)]{kol17} Kollmeier, J. A., Zasowski, G., Rix, H. W., et al. 2017,  arXiv: astro-ph/171103234 
\bibitem[Kormendy et al. (2011)]{kor11} Kormendy, J., Bender, R., \& Cornell, M. E. 2011, \nat, 469, 374
\bibitem[Kriss (1994)]{kri94} Kriss, G. 1994, in ASP Conf. Ser. 61, Astronomical Data Analysis Software
and Systems III, ed. D. R. Crabtree, R. J. Hanisch, \& J. Barnes (San Fransisco, CA: ASP), 437
\bibitem[Krolik (1999)]{kro99} Krolik, J. H. 1999, Active Galactic Nuclei: From the Central Black Hole to the Galactic Environment (Princeton University Press: Princeton, NJ)
\bibitem[LaMassa et al. (2015)]{lam15} LaMassa, S. M., Cales, S., Moran, E. C., et al. 2015, \apj, 800, 144
\bibitem[Lawrence (2018)]{law18} Lawrence, A. 2018, \nat\ Astronomy, 2, 102
\bibitem[Liu et al. (2021)]{liu21} Liu, W. J., Lira, P., Yao, S., Xu, D. W., Wang, J., Dong, X. B., \& Martínez-Palomera, J. 2021, \apj, 915, 63
\bibitem[Lodato \& Rossi (2011)]{lor11} Lodato, G., \& Rossi, E. M. 2011, \mnras, 410, 359
\bibitem[L{\'o}pez-Navas et al.(2022)]{lop22} L{\'o}pez-Navas, E., Mart{\'\i}nez-Aldama, M.~L., Bernal, S., et al.\ 2022, \mnras, 513, L57
\bibitem[L{\'o}pez-Navas et al.(2023)]{lop23} L{\'o}pez-Navas, E., S{\'a}nchez-S{\'a}ez, P., Ar{\'e}valo, P., et al.\ 2023, \mnras, 524, 188
\bibitem[Lyke et l. (2020)]{lyk20} Lyke, B. W., Higley, A. N., McLane, J. N., et al. 2020, \apjs, 250., 8 
\bibitem[Lyu et al. (2022)]{lyu22} Lyu, B., Wu, Q., Yan, Z., et al. 2022, \apj, 927, 227
\bibitem[MacLeod et al. (2019)]{mac19} MacLeod, C. L., Green, P. J., Anderson, S. F., et al. 2019, \apj, 874, 8
\bibitem[MacLeod et al. (2010)]{mac10} MacLeod, C. L., Ivezic, Z., Kochanek, C. S., et al. 2010, \apj, 721, 1014 
\bibitem[MacLeod et al. (2016)]{mac16} MacLeod, C. L., Ross, N. P., Lawrence, A., et al. 2016, \mnras, 457, 389
\bibitem[Mainzer et al. (2014)]{mai14} Mainzer, A., Bauer, J., Cutri, R. M., et al. 2014, \apj, 792, 30
\bibitem[Marin et al. (2019)]{mar19} Marin, F., Hutsemekers, D., \& Agis Gonzalez, B. 2019, proceedings of the 2019's annual conference of the SF2A , arXiv/astro-ph:1909.02801
\bibitem[Malkan \& Sargent (1982)]{mas82} Malkan, M. A., \& Sargent, W. L. W. 1982, \apj, 254, 22
\bibitem[Marinucci et al. (2016)]{mar16} Marinucci, A., Bianchi, S., Nicastro, F., Matt, G., \& Goulding, A. D. 2012, \apj, 748, 130
\bibitem[Marziani \& Sulentic (2012)]{mas12} Marziani, P., \& Sulentic, J. W. 2012, NewAR, 56, 49
\bibitem[Massey et al. (1988)]{mas88} Massey, P., Strobel, K., Barnes, J. V., et al. 1988, \apj, 328, 315
\bibitem[Mathur (2000)]{mat00} Mathur, S. 2000, \mnras, 314, 17
\bibitem[Mathur et al. (2018)]{mat18} Mathur, S., Denney, K. D., Gupta, A., et al. 2018, \apj, 886, 123
\bibitem[Mathur et al. (2012)]{mat12} Mathur, S., Fields, D., Peterson, B. M., \& Grupe, D. 2012, \apj, 754, 146  
\bibitem[McElroy et al. (2016)]{mc16} McElroy, R. E., Husemann, B., Croom, S. M., et al. 2016, \aap, 593, L8
\bibitem[Merloni et al. (2015)]{mer15}  Merloni, A., Dwelly, T., Salvato, M.,  et al. 2015, \mnras, 452, 69 
\bibitem[Miller \& Stone (1994)]{mis94} Miller, J. S., \& Stone, R. P. S. 1994, Lick Obs. Tech. Rep. 66 (Santa Cruz, CA: Lick Observatory) 
\bibitem[Miniutti et al. (2019)]{min19} Miniutti, G., Saxton, R. D., Giustini, M., et al. 2019, \nat, 573, 381
\bibitem[Nagoshi et al. (2021)]{nag21}  Nagoshi, S., Iwamuro, F., Wada, K., \& Saito, T. 2021, \pasj, 73, 122
\bibitem[Newman et al. (2013)]{new13} Newman, J. A., Cooper, M. C., Davis, M., et al. 2013, \apjs, 208, 5 
\bibitem[Netzer (2013)]{zet13} Netzer, H. 2013, The Physics and Evolution of Active Galactic Nuclei, by Hagai Netzer, Cambridge, UK: Cambridge University Press, 2013
\bibitem[Nicastro (2000)]{nic00} Nicastro, F. 2000, \apjl, 530, 65
\bibitem[Oke et al. (1995)]{oke95} Oke, J. B., Cohen, J. G., Carr, M., et al. 1995, \pasp, 107, 375 
\bibitem[Orban de Xivry et al. (2011)]{orb11} Orban de Xivry, G., Davies, R., Schartmann, M., Komossa, S., Marconi, A., Hicks, E., Engel, H., \& Tacconi, L. 2011, \mnras, 417, 2721
\bibitem[Pan et al. (2021)]{pan21} Pan, X., Li, S. -L., \& Cao, X. 2021, arXiv:astro-ph/2103:00828, accetped by \apj
\bibitem[Parker et al. (2016)]{par16} Parker, M. L., Komossa, S., Kollatschny, W., et al. 2016, \mnras, 461, 1927
\bibitem[Parker et al. (2019)]{par19} Parker, M. L., Schartel, N., Grupe, D., et al. 2019, \mnras, 483, L88
\bibitem[Perez-Montero \& Diaz (2003)]{ped03} Perez-Montero, E., \& Diaz, A. I. 2013, \mnras, 346, 105
\bibitem[Perley (2019)]{per19} Perley, D. A. 2019, \pasp, 131h4503P
\bibitem[Peterson (2014)]{pet14} Peterson, B. M. 2014, SSRv, 183, 253 
\bibitem[Peterson \& Bentz (2006)]{peb} Peterson, B. M., \& Bentz, M. C. 2006, NewAR, 50, 769
\bibitem[Peterson et al. (2000)]{pet00} Peterson, B. M., McHardy, I. M., Wilkes, B. J., et al. 2000, \apj, 542, 161
\bibitem[Pizzolato \& Soker (2005)]{pis05} Pizzolato, F., \& Soker, N. 2005, \apj, 632, 821
\bibitem[Rees (1988)]{ree88} Rees, M. J. 1988, \nat, 333, 523
\bibitem[Ricci \& Trakhtenbrot (2022)] {rit22}  Ricci, C., \& Trakhtenbrot, B. 2022,  arXiv: astro-ph/221105132
\bibitem[Risaliti et al. (2009)]{ris09} Risaliti, G., Salvati, M., Elvis, M., et al. 2009, \mnras, 393, L1
\bibitem[Ross et al. (2018)]{ros18} Ross, N. P., Ford, K. E. S., Graham, M., et al. 2018, \mnras, 480, 4468 
\bibitem[Ruan et al. (2016)]{rua16} Ruan, J. J., Anderson, S. F., Cales, S. L., et al. 2016, \apj, 826, 188
\bibitem[Ruan et al. (2019)]{rua19} Ruan, J. J., Anderson, S. F., Eracleous, M., Green, P. J., Haggard, D., MacLeod, C. L.\, Runnoe, J. C., \& Sobolewska, M. A. 2019, \apj, 883, 76 
\bibitem[Runnoe et al. (2016)]{tun16} Runnoe, J. C., Cales, S., Ruan, J. J., et al. 2016, \mnras, 455, 1691
\bibitem[Sani et al. (2010)]{san10} Sani, E., Lutz, D., Risaliti, G., et al. 2010, \mnras, 403, 124
\bibitem[Shakura \& Sunyaev (1973)]{shs73} Shakura, N. I., \& Sunyaev, R. A. 1973, \aap, 24, 337
\bibitem[Shapovalova et al. (2010)]{sha10} Shapovalova, A. I., Popovic, L. C., Burenkov, A. N., et al. 2010, \aap, 509, 106
\bibitem[Shappee et al. (2014]{sha14} Shappee, B. J., Prieto, J. L., Grupe, D., et al. 2014, \apj, 788, 48
\bibitem[Scharwachter et al. (2017)]{sch17} Scharwachter, J., Husemann, B., Busch, G., Komossa, S., \& Dopita, M. A. 2017, \apj, 848, 35
\bibitem[Schlegel \& Finkbeiner (2011)]{scf11} Schlafly, E. F., \& Finkbeiner, D. P. 2011, \apj, 737, 103
\bibitem[Schlegel et al. (1998)]{sch98} Schlegel, D., Finkbeiner, D. P., \& Davis, M. 1998, \apj, 500, 525
\bibitem[Shen et al. (2011)]{shen11} Shen, Y., Richards, G. T., Strauss, M. A., et al. 2011, \apjs, 194, 45
\bibitem[Sheng et al. (2017)]{she17} Sheng, Z., Wang, T., Jiang, N., et al. 2017, \apjl, 846, 7
\bibitem[Sheng et al. (2020)]{she20} Sheng, Z., Wang, T., Jiang, N., et al. 2020, \apj, 889, 46  
\bibitem[Siemiginowska et al. (1996)]{sie96} Siemiginowska, A., Czerny, B., \& Kostyunin, V. 1996, \apj, 458, 491
\bibitem[Smee et al. (2013)]{sme13} Smee, S. A., Gunn, J. E., Uomoto, A., et al. 2013, \aj, 146, 32  
\bibitem[Sniegowska et al. (2020)]{sni00} Sniegowska, M., Czerny, B., Bon, E., \& Bon, N. 2020, \aap, 641, 167  
\bibitem[Stern et al. (2018)]{ste18} Stern, D., McKernan, B., Graham, M. J., et al. 2018, \apj, 864, 27
\bibitem[Storchi-Bergmann et al. (2017)]{sto17} Storchi-Bergmann, T., Schimoia, J. S., Peterson, B. M., Elvis, M., Denney, K. D., Eracleous, M., \& Nemmen, R. S. 2017, \apj, 835, 236
\bibitem[Storey \& Hummer(1995)]{sth95} Storey, P. J., \& Hummer, D. G. 1995, \mnras, 272, 41
\bibitem[Strateva et al. (2003))]{atr03} Strateva, I. V., Strauss, M. A., Hao, L., et al. 2003, \aj, 126, 1720
\bibitem[Tody (1986)]{tod86} Tody, D. 1986, Proc. SPIE, 627, 733
\bibitem[Tody (1992)]{tod92} Tody, D. 1992, in ASP Conf. Ser. 52, Astronomical Data Analysis Software
and Systems II, ed. R. J. Hanisch, R. J. V. Brissenden, \& J. Barnes (San Fransisco, CA: ASP), 173
\bibitem[Trakhtenbrot et al. (2019)]{tra19} Trakhtenbrot, B., Arcavi, I., MacLeod, C. L., et al. 2019, \apj, 883, 94
\bibitem[van Velzen \& Farrar (2014)]{vaf14} van Velzen, S., \& Farrar, G. R. 2014, \apj, 792, 53 
\bibitem[Veron-Cetty et al. (2004)]{ver04} Veron-Cetty, M. -P., Joly, M., \& Veron, P. 2004, \aap, 417, 515
\bibitem[Voges et al. (1999)]{vog99} Voges, W., Aschenbach, B., Boller, T., et al. 1999,  \aap, 349, 389
\bibitem[Wang (2015)]{wan15} Wang, J. 2015, NewA, 37, 15
\bibitem[Wang et al. (2011)]{wan11} Wang, J., Mao, Y. F., \& Wei, J. Y. 2011, \apj, 741, 50
\bibitem[Wang \& Wei (2008]{waw08} Wang, J., \& Wei, J. Y. 2008, \apj, 679, 86
\bibitem[Wang \& Wei (2010)]{waw10} Wang, J., \& Wei, J. Y. 2010, \apj, 719, 1157
\bibitem[Wang et al. (2006)]{wan06} Wang, J., Wei, J. Y., \& He, X. T. 2006, \apj, 638, 106
\bibitem[Wang et al. (2020a)]{wan20a} Wang, J., Xu, D. W., Sun, S. S., Feng, Q. C. Li T. R., Xiao, P. F., \& Wei, J. Y. 2020a, \aj, 159, 245
\bibitem[Wang et al. (2020b)]{wan20b} Wang, J., Xu, D. W., \& Wei, J. Y. 2020b, \apj, 901, 1
\bibitem[Wang et al. (2018)]{wan18} Wang, J., Xu, D. W., \& Wei, J. Y. 2018, \apj, 852, 26
\bibitem[Wang et al. (2019)]{wan19} Wang, J., Xu, D. W., Wang, Y., Zhang, J. B., Zheng, J., \& Wei, J. Y. 2019, \apj, 887, 15
\bibitem[Wang et al. (2018)]{wan18} Wang, J., Xu, D. W., \& Wei, J. Y. 2018, \apj, 858, 49 
\bibitem[Wang et al. (2022a)]{wan22a} Wang, J., Zheng, W. K., Xu, D. W.,  Brink, T. G., Filippenko, A. V., Gao, C., Sun, S. S., \& Wei, J. Y. 2022, RAA, 22, 015011
\bibitem[Wang et al. (2013)]{wan13} Wang, J., Zhou, X. L.,  \& Wei, J. Y. 2013, \apj, 768, 176
\bibitem[Wang et al. (2014)]{wan14} Wang, J. M., Du, P., Hu, C., et al. 2014, \apj, 793, 108
\bibitem[Wang \& Bon (2020)]{wab20} Wang, J. -M., \& Bon, E. 2020, \apjl, 643, 9
\bibitem[Wang et la. (2022b)]{wan22b} Wang, Y., Jiang, N., Wang, T., et al. 2022b, \apjs, 258, 21
\bibitem[Wild et al. (2010)]{wil10} Wild, V., Heckman, T. M., \& Charlot, S. 2010, \mnras, 405, 933
\bibitem[Wild et al. (2007)]{wil07} Wild, V., Kauffmann, G., Heckman, T., et al. 2007, \mnras, 381, 543
\bibitem[Winter et al. (2012)]{win12} Winter, L. M., Veilleux, S., McKernan, B., \& Kallman, T. R. 2012, \apj, 745, 107
\bibitem[Woo et al. (2017)]{woo17} Woo, J. H., Son, D., \& Bae, H. J. 2017, \apj, 839, 120
\bibitem[Worthey \& Ottaviani (1997)]{woo97} Worthey, G., \& Ottaviani, D. L. 1997, \apjs, 111, 377
\bibitem[Wright et al. (2010)]{wri10} Wright, E. L., Eisenhardt, P. R. M., Mainzer, A. K., et al. 2010, \aj, 140,
1868
\bibitem[Wu et al. (2004)]{wuw04} Wu, X. B., Wang, R., Kong, M. Z., Liu, F. K., \& Han, J. L. 2004, \aap, 424, 793
\bibitem[XMM-SSC (2018)]{xmm18} XMM-SSC 2018, VizieR Online Data Catalog: XMM-Newton Slew Survey
Source Catalogue, version 2.0
\bibitem[Yan et al. (2019)]{yan19} Yan, L., Wang, T. G., Jiang, N., et al. 2019, \apj, 874, 44
\bibitem[Yang et al. (2018)]{yan18} Yang, Q., Wu, X. B., Fan, X. H., et al. 2018, \apj, 862, 109
\bibitem[Yu et al. (2020)]{yu20} Yu, X., Shi, Y., Chen, Y., et al. 2020, \mnras, 498, 3985
\bibitem[Zhang et al. (2013)]{zha13} Zhang, K., Wang, T., Gaskell, C. M., \& Dong, X. 2013, \apj, 762, 51
\bibitem[Zhou et al. (2006)]{zho06} Zhou, H. Y., Wang, T. G., Yuan, W. M., et al. 2006, \apjs, 166, 128
\bibitem[Zimmermann et al. (2001)]{zim01} Zimmermann, H. -U., Boller, T., Döbereiner, S., \& Pietsch, W. 2001, \aap, 378, 30


\end{thebibliography}

\end{document}